%% file: main.tex
\documentclass[final,5p,authoryear,twocolumn]{elsarticle}

\usepackage{amssymb}
\usepackage{amsmath}
\usepackage{svg}
\usepackage{tikz}
\usepackage{calc}
\usetikzlibrary{decorations.markings}
\usepackage{bm}
\usepackage{acronym}                      
\usepackage{siunitx}
\usetikzlibrary{calc,arrows.meta,decorations.pathreplacing,positioning,shapes.geometric}
\usepackage[dvipsnames]{xcolor}
\newcommand{\mathsfbi}[1]{\bm{\mathsf{#1}}}
\usepackage{tikz-cd}

\definecolor{bb22}{HTML}{f54952}
\definecolor{bb14}{HTML}{f78c6b}
\definecolor{bu0}{HTML}{999999}
\definecolor{bd0}{HTML}{4D4D4D}
\definecolor{bt14}{HTML}{00A37D}
\definecolor{bt22}{HTML}{002EA3}

\DeclareSIUnit\px{px}

\acrodef{PIV}[PIV]{Particle Image Velocimetry}
\acrodef{ROM}[ROM]{Reduced-Order Model}
\acrodefplural{ROM}[ROMs]{Reduced-Order Models}
\acrodef{PTV}[PTV]{Particle Tracking Velocimetry}
\acrodef{FOV}[FOV]{Field Of View}
\acrodef{URANS}[URANS]{Unsteady Reynolds-Averaged Navier-Stoker}
\acrodef{PVC}[PVC]{Polyvinyl Chloride}
\acrodef{POD}[POD]{Proper Orthogonal Decomposition}
\acrodef{LIC}[LIC]{Line Integral Convolution}

\journal{Experimental Thermal and Fluid Science}

\begin{document}

\begin{frontmatter}



\title{On the turbulent wake of the actuated fluidic pinball: dynamics, bifurcations and control authority}


\author[UC3M]{Alicia Rodríguez-Asensio} 
\author[UC3M]{Luigi Marra}
\author[UC3M]{Ignacio Andreu-Angulo}
\author[UC3M_st]{Andrea Meilán-Vila}
\author[INTA,UC3M]{Juan Alfaro Moreno}
\author[UC3M]{Guy~Y.~Cornejo~Maceda}
\author[Shenzhen_2,Shenzhen_1,UC3M]{Bernd R. Noack}
\author[UC3M]{Andrea Ianiro}
\author[UC3M]{Stefano Discetti}


\affiliation[UC3M]{organization={Department of Aerospace Engineering, Universidad Carlos III de Madrid},
            addressline={Avenida Universidad, 30}, 
            postcode={28911}, 
            state={Leganés},
            country={Spain}}
\affiliation[UC3M_st]{organization={Department of Statistics, Universidad Carlos III de Madrid},
            addressline={Avenida Universidad, 30}, 
            postcode={28911}, 
            state={Leganés},
            country={Spain}}
            
\affiliation[INTA]{organization={Aerial Platforms Department, Spanish National Institute for Aerospace Technology (INTA)},
            addressline={Ctra. M-301, Km 10, 500}, 
            postcode={28830}, 
            state={San Martín de la Vega},
            country={Spain}}

\affiliation[Shenzhen_2]{organization={College of Mechatronics and Control Engineering, Shenzhen University},
addressline={Canghai campus},
postcode={518060},
state={Shenzhen},
country={PR China}}
\affiliation[Shenzhen_1]{organization={Guangdong Province VTOL Aircraft Manufacturing Innovation Center},
postcode={518060},
state={Shenzhen},
country={PR China}}

\begin{abstract}
\input{Chapters/0-Abstract}

\end{abstract}




\end{frontmatter}



\input{Chapters/1-Introduction}

\input{Chapters/2-Methodology}

\input{Chapters/3-Results}

\input{Chapters/4-ROM_discussion}

\input{Chapters/5-Conclusions}

\section*{Acknowledgments}
This work was mainly supported by the project EXCALIBUR (Grant No. PID2022-138314NB-468 I00), funded by MCIU/AEI/10.13039/501100011033 and by “ERDF A way of making Europe”. LM was funded by the “Orden 3789/470 2022, del Vicepresidente, Consejero de Educación y Universidades, por la que se convocan ayudas para la contratación de personal investigador predoctoral en formación para el año 2022”. The work was also partially supported by the research project PREDATOR-CM-UC3M. This project has been funded by the call "Estímulo a la Investigación de Jóvenes Doctores/as" within the frame of the Convenio Plurianual CM-UC3M and the V PRICIT (V Regional Plan for Scientific Research and Technological Innovation). This work is also supported by the National Science Foundation of China (NSFC) 
through grants 
 W2541002, 
and by the Shenzhen Science and Technology Innovation Program 
under grants KJZD20230923115210021  
and JCYJ20220531095605012. 

\section*{CRediT authorship contribution statement}
\noindent \textbf{ARA}: Data curation, Formal analysis, Investigation, Methodology, Software, Validation, Visualization, Writing -- original draft, Writing -- review \& editing. \textbf{LM}: Investigation, Methodology, Writing -- review \& editing. \textbf{IAA}: Investigation, Methodology, Software, Visualization, Writing -- original draft, Writing -- review \& editing. \textbf{AMV}: Conceptualization, Funding acquisition, Methodology, Project administration, Supervision, Writing -- review \& editing. \textbf{JAM}: Methodology, Software, Writing -- review \& editing. \textbf{GYCM}: Methodology, Supervision, Writing -- review \& editing. \textbf{BRN}: Conceptualization, Methodology, Resources, Supervision, Writing -- review \& editing. \textbf{AI}: Conceptualization, Formal analysis, Funding acquisition, Methodology, Project administration, Resources, Supervision, Writing -- review \& editing. \textbf{SD}: Conceptualization, Formal analysis, Funding acquisition, Methodology, Project administration, Resources, Supervision, Writing -- review \& editing.

\section*{Declaration of generative AI in scientific writing}
During the preparation of this work, the authors used generative Artificial Intelligence in the writing process to improve the readability and language of the manuscript. After using this tool, the authors reviewed and edited the content as needed and assume full responsibility for the content of the published article.


\appendix
\input{Chapters/A-Uncertainty}







\end{document}

%% file: Chapters/0-Abstract.tex
We present the first comprehensive experimental and numerical study 
featuring the turbulent wake of the fluidic pinball
for a large actuation range. The fluidic pinball  is a cluster of three equal circular cylinders
centered on the vertices of an equilateral triangle, pointing upstream in uniform flow.
This configuration has become a canonical benchmark 
for control-oriented reduced-order modeling,
for nonlinear control design and 
for a large kaleidoscope of drag reduction mechanisms.
While the literature covers well the laminar two-dimensional Reynolds number regime,
 we focus on unexplored terra incognita:
 experiments of the symmetrically actuated turbulent regime 
 at a Reynolds number of $Re=9100$.
 In other words, the upstream cylinder is kept stationary, 
while the two downstream cylinders rotate with equal and opposite angular velocities.
A large range of base-bleeding and boat-tailing actuation parameters
 is investigated with time-resolved particle image velocimetry
and aerodynamic force measurement with a companion Reynolds-averaged Navier-Stokes simulation.
Our results indicate that the turbulent wake of the fluidic pinball 
can be approximated by a three-dimensional actuation manifold \citep{marra2024actuation} 
comprising two inverse pitchfork bifurcations.
In the boat-tailing limit, a reduced control authority with a new low-frequency shedding state is observed.

%% file: Chapters/1-Introduction.tex
\section{Introduction}
\label{sec:introduction}
We experimentally investigate the turbulent wake dynamics of the actuated fluidic pinball under symmetric rotation of the downstream cylinders. The fluidic pinball is a benchmark configuration for controlled bluff-body wake flows and reduced-order modeling, yet its turbulent-regime behavior remains poorly understood. A key open question is whether wake transitions and low-dimensional organization induced by actuation, as those observed by \cite{deng2020low}, persist at higher Reynolds numbers ($Re$). Here, we address this question using synchronized time-resolved velocity field and force measurements, complemented by numerical simulations.

Multiple cylinder-like structures are present in many engineering applications, ranging from heat exchangers and nuclear power plant cooling systems to offshore structures, buildings, power lines, and cables. Flows around multiple bluff bodies feature strong wake interactions and characteristic vortex patterns. Understanding and manipulating these wake dynamics is thus paramount for a wide range of applications. 
Numerous studies have examined flows over configurations involving multiple cylinders, including two side-by-side cylinders \citep{sumner2000flow,sumner2010two}, three cylinders forming an equilateral triangle \citep{lam1988phenomena,bansal2017experimental,raibaudo2020machine,Chen2020jfm,raibaudo2021unsteady}, and larger clusters \citep{lam2003force}. Many of these works show that the wake dynamics depend on geometric parameters, such as the angle of attack and the spacing between  the cylinders, but are mainly governed by the Reynolds number. 

\begin{table*}[t!]
\centering

\input{Figures/Intro_table}

\caption{Representative studies on multi-cylinder wakes and the fluidic pinball, highlighting configuration, Reynolds-number range, actuation strategy, and measurement approach. They are separated depending on whether experiments were performed, simulations were used, and if forces were directly measured.}
\label{tab:literature_comparison}
\end{table*}


For configurations involving two circular cylinders, wake dynamics depend on the relative arrangement and Reynolds number. \citet{sumner2010two} documented the wide range of flow regimes observed for tandem, staggered, and side-by-side cylinder configurations in both laminar and turbulent flows. For the side-by-side configuration at $Re=2100$, a biased gap-flow regime commonly develops, in which the jet between the cylinders is deflected toward one cylinder, leading to asymmetric wakes. Depending on the spacing, this configuration may exhibit a bi-stable or quasi-stable behavior, with intermittent or forced switching of the gap-flow direction. The occurrence and persistence of these states are sensitive to external perturbation, inflow conditions, and control inputs. These features foreshadow the even richer multistability observed in three-cylinder configurations.


The wake dynamics of clusters of three circular cylinders in an equilateral configuration are conditioned by the Reynolds number and their orientation with respect to the incoming flow. As shown by \citet{lam1988phenomena}, a dominant gap jet between the two downstream cylinders is reported, and its orientation is subject to the initial conditions and the geometric alignment of the cluster. The wake settles into one of two quasi-stable states, with no spontaneous or intermittent switching observed in the absence of external perturbations. 
This sensitivity is particularly pronounced in a specific orientation known as the fluidic pinball, in which one side of the equilateral triangle is perpendicular to the incoming flow, and the opposite vertex is pointing upstream. \citet{deng2020low} identified a sequence of bifurcations as the Reynolds number increases. The steady symmetric wake is stable up to $Re \approx 18$ (based on the incoming velocity and the cylinder diameter), beyond which a Hopf bifurcation gives rise to periodic vortex shedding and the formation of a gap jet between the downstream cylinders. A subsequent pitchfork bifurcation at $Re \approx 68$ destabilizes the symmetric state, leading to two mirror-conjugate asymmetric solutions characterized by upward or downward gap-jet deflection. In all cases, these steady solutions are unstable to von Kármán shedding in the post-bifurcation regime. 

The interest in controlling vortex shedding in many engineering applications (e.g., reduction of vortex-induced vibrations and mitigation of unsteady loads on structures) has motivated the development of several solutions. Passive methods, such as surface roughness \citep{achenbach1971influence} or splitter plates \citep{apelt1973effects,kwon1996control}, can effectively stabilize the wake, but their performance is generally restricted to a specific flow configuration, \citep[e.g., fixed Reynolds number or geometric conditions as shown by][]{kwon1996control}. In contrast, active control provides flexibility, as it can adapt to changing conditions and can also be switched off on demand \citep{Brunton2015amr}. A common example is cylinder rotation, which modifies the motion of the boundary layer and alters the wake response~\citep{tokumaru1991rotary}. 
These findings motivated the use of controlled rotation in multi-body configurations. 
The fluidic pinball with independently rotating cylinders is capable of reproducing up to six canonical wake-manipulation mechanisms \citep{maceda2021stabilization}. In particular, boat tailing and base bleeding are the two control actuation mechanisms with the strongest influence on drag.
Boat tailing vectors the flow towards the center and narrows the wake, whereas base bleeding enhances the gap jet and prevents the development of a vortex near the cylinders.

Table \ref{tab:literature_comparison} summarizes representative studies on multi-cylinder configurations, highlighting the Reynolds number range, actuation strategy, and methodology. Although previous studies of the fluidic pinball have provided detailed flow velocity measurements at $Re<2500$ \citep{bansal2017experimental,raibaudo2020machine, raibaudo2021unsteady}, they have focused on the wake kinematics without synchronized force measurements. 
They generally report monotonic trends toward improved wake symmetry and drag reduction with increasing boat tailing actuation. However, the literature data on the combined effect of controlled rotation on both the wake dynamics and the resulting aerodynamic forces is limited, especially in the turbulent regime. 

To address this gap, we perform time-resolved \ac{PIV} alongside synchronized force measurements on an actuated fluidic pinball at $Re=9100$. We cover a wide range of rotational speeds, imposing counter-rotation of the two downstream cylinders. These measurements enable a detailed characterization of the wake dynamics, modal structure, and effect of the actuation on the aerodynamic forces.
The experimental findings are complemented with \ac{URANS} simulations. It is acknowledged that \ac{URANS} has limitations in accurately predicting flow separation and wake dynamics \citep{Nishino2008}, but it can capture key physics of the flow around smooth bodies such as cylinders \citep{ding2013}. While the numerical effort is not intended to replace the experimental analysis, it provides insight into the near-cylinder and inter-cylinder regions that are difficult to resolve experimentally.

The manuscript is organized as follows. The methodology for experimental data processing and simulations, together with the experimental setup, is described in Section \ref{sec:methodology}. Flow field and force measurements, along with numerical results, are presented in Section \ref{sec:results}. Section \ref{sec:ROM} summarizes the results with a reduced-order representation. Finally, Section \ref{sec:conclusions} provides a discussion of the results and presents the conclusions.

%% file: Figures/Intro_table.tex
\begin{tabular}{llllccc}
\hline
Reference & Configuration & $Re$ & Control & Exp. & CFD & Forces \\
\hline 
\citet{achenbach1971influence} 
& 1 cyl.
& $3\times10^6$ 
& Roughness 
& \checkmark 
& - 
& \checkmark \\
\citet{apelt1973effects} 
& 1 cyl.
& $10^4$--$5\times10^4$ 
& Splitter plate
& \checkmark 
& - 
& \checkmark \\
\citet{lam1988phenomena} 
& 3 cyl., triangular 
& $2100,\,3500$ 
& - 
& \checkmark 
& - 
& \checkmark \\
\citet{tokumaru1991rotary} 
& 1 cyl.
& $15000$
& Cyl. rotation 
& \checkmark 
& - 
& \checkmark \\
\citet{kwon1996control} 
& 1 cyl.
& $\sim 10^4$ 
& Splitter plate
& \checkmark 
& - 
& \checkmark \\

\citet{sumner2000flow} 
& 2 cyl., staggered 
& $850$--$1900$ 
& -
& \checkmark  
& - 
& - \\

\citet{lam2003force} 
& 4 cyl.
& $\sim 10^4$ 
& -
& \checkmark 
& -
& - 
\\

\citet{sumner2010two} 
& 2 cyl. multiple
& $10^2$--$10^5$ 
& -
& \checkmark  
& \checkmark 
& \checkmark \\

\citet{bansal2017experimental} 
& 3 cyl., triangular 
& $2100$ 
& -
& \checkmark  
& - 
& - \\

\citet{Chen2020jfm} 
& 3 cyl., triangular 
& $50$--$300$ 
& -
& -   
& \checkmark 
& \checkmark \\

\citet{deng2020low} 
& Fluidic pinball 
& $\leq 130$ 
& -
& -
& \checkmark
& \checkmark \\

\citet{raibaudo2020machine} 
& Fluidic pinball 
& $2200$ 
& Cyl. rotation
& \checkmark
& - 
& - \\

\citet{raibaudo2021unsteady} 
& Fluidic pinball 
& $2200$ 
& Cyl. rotation
& \checkmark
& - 
& - \\

\citet{maceda2021stabilization} 
& Fluidic pinball 
& $100$ 
& Cyl. rotation
& - 
& \checkmark
& - \\

Present work 
& Fluidic pinball 
& $9100$ 
& Cyl. rotation
& \checkmark
& \checkmark
& \checkmark \\
\hline
\end{tabular}

%% file: Chapters/2-Methodology.tex
\section{Methodology}
\label{sec:methodology}

This section describes the experimental setup, measurement, and numerical techniques employed in the present study. First, the experimental arrangement and actuation strategy are detailed, including the fluidic pinball configuration, operating conditions, and control parameters. The velocity field and force measurement procedures are then described, together with data processing, calibration, and uncertainty estimation. Finally, the numerical methodology of the \ac{URANS} simulations is presented.

\subsection{Experimental setup}
\label{subsec:exp_setup}
The experiments were performed in a closed-loop water tunnel at Universidad Carlos III de Madrid. The test section is $\SI{2.5}{\metre}$ long with a rectangular cross-section of $\SI{0.5}{\metre}$ by $\SI{0.55}{\metre}$. The sidewalls are made of glass to provide full optical access. The freestream velocity can be set between $\SI{0.1}{\metre\per\second}$ and $\SI{2}{\metre\per\second}$. 

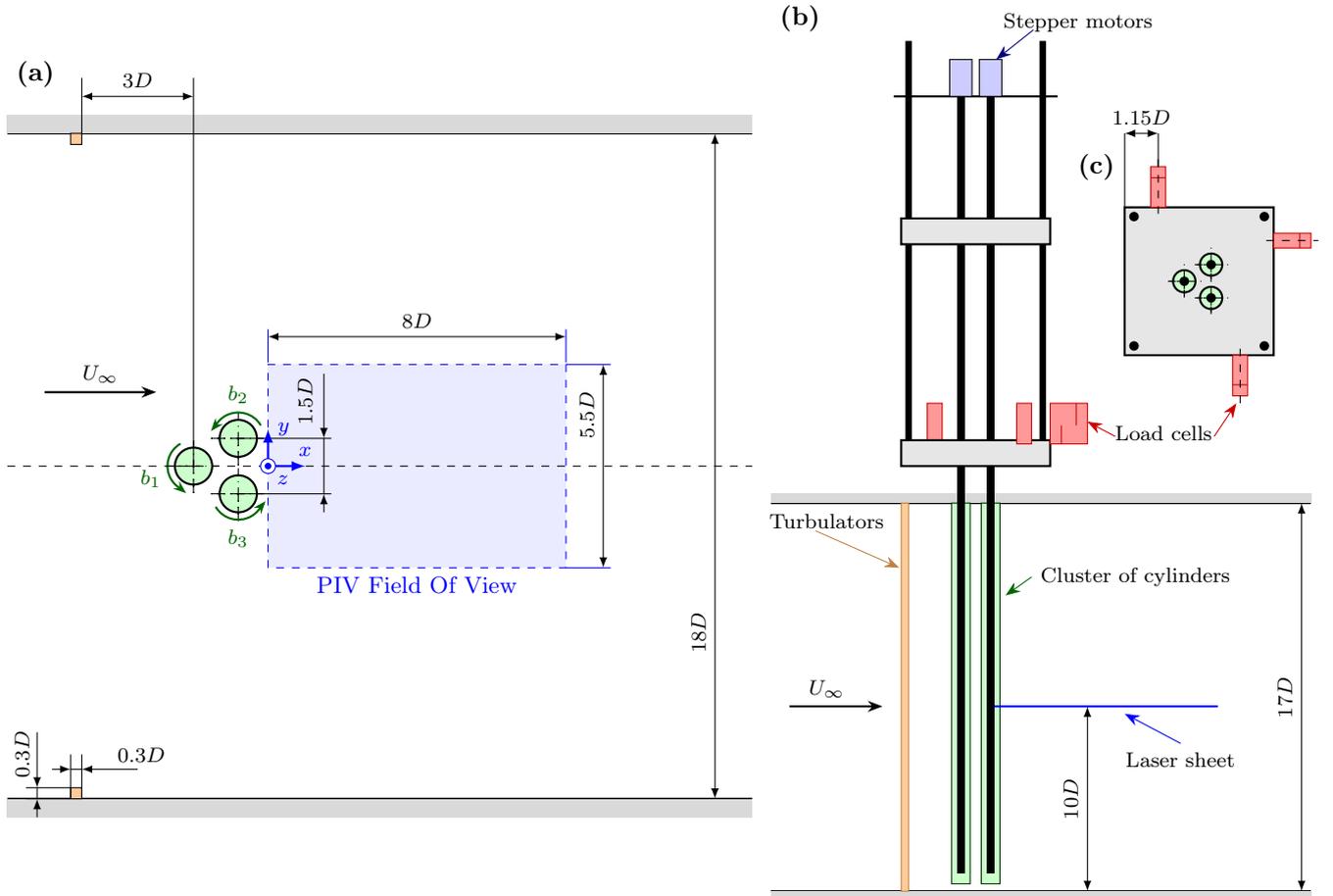
\begin{figure*}[ht]
\centering
\input{Figures/Flow_config}
\caption{Schematics of the experimental setup. (a) Two-dimensional representation of the fluidic pinball with main dimensions and nomenclature. The blue area depicts the PIV measurement domain. The brown squares represent the turbulators placed upstream of the cylinders. (b) Side view of the arrangement. (c) Top view of the lower plate where the load cells are mounted.}
\label{fig:2PIV_setup}
\end{figure*}

Figure \ref{fig:2PIV_setup}(a) shows a two-dimensional sketch of the fluidic pinball installed in the water tunnel. It consists of three circular cylinders of diameter $D=\SI{30}{\milli\metre}$, whose axes are orthogonal to the freestream. 
The axes of the cylinders are equally spaced $1.5D$ center-to-center, forming an equilateral triangle. This results in a geometrical blockage of approximately $14\%$. The upstream cylinder is labeled \textit{1}, while the downstream ones are \textit{2} and \textit{3}. Their actuation parameters (e.g., rotation speeds $\Omega_i,\ i = 1,2,3$) are identified using the corresponding subscripts \textit{1}, \textit{2}, and \textit{3}, respectively.
The cylinders were machined in plexiglass and reinforced internally with $\SI{20}{\milli\metre}$ rectified steel bars to avoid undesirable eccentric rotations. 
The experimental setup installed in the test section is presented in Figure \ref{fig:2PIV_setup}(b).
The cylinders were fixed to two \ac{PVC} square plates and one aluminum square plate, connected with stiff threaded bars. At each cylinder-bar interface, an NKI20/20-XL needle roller bearing ensured axis alignment. In the upper \ac{PVC} plate, one SKF 51204 single-direction thrust ball bearing per cylinder guarantees smooth rotation. The entire assembly was suspended from a top beam structure using metallic wire, allowing precise alignment with the incoming flow. 
The cylinders were supported and controlled only at their upper ends, which were mechanically connected to the load-cell assembly. The lower ends were left mechanically unconstrained, allowing the entire cluster to transmit hydrodynamic loads to the load cell without over-constraint.

The experiments were conducted at a freestream velocity $U_\infty= \SI{31}{\centi\metre\per\second}$, corresponding to a Reynolds number $Re=U_\infty D/\nu=9100$, where $\nu=\SI{1}{\milli\metre\squared\per\second}$ is the kinematic viscosity of water. 
This Reynolds number is more than 4 times larger than the highest value tested in other fluidic pinball experiments (see Table \ref{tab:literature_comparison}). The upper limit for the freestream velocity was set by the requirements of time-resolved \ac{PIV} with the available hardware.
The incoming-flow turbulence intensity at this free-stream velocity is approximately $2\%$.
Turbulators were placed on each lateral wall of the water tunnel upstream of the pinball to promote transition to a turbulent incoming flow. This was intended to avoid the formation of a laminar separation bubble on the lateral walls due to the adverse pressure gradient experienced downstream of the pinball.

Each cylinder was driven independently at one end by a SY28STH45-0674A stepper motor located above the upper aluminum plate. The motors were driven using Tic T825 multi-interface controllers connected to an Arduino Mega 2560. 
The resulting non-dimensional rotational speed of the cylinders is expressed as:
\begin{equation}
    b_i=\frac{\Omega_i D}{2U_\infty},\ i=1,2,3.
\end{equation}
This parameter represents the ratio between the tangential velocity of the cylinder surface and the freestream velocity. Since $b_i$ depends on $U_\infty$, the selection of the operating Reynolds number also constrains the accessible actuation range. The maximum achievable angular velocity of the motors was limited not only by their torque capacity but also by the inertia and weight of the supported assembly. 
The Arduino Mega 2560 was also employed to synchronize the actuation with the velocity and force measurements, triggering the load-cell acquisition system, the laser for illuminating the seeding particles, and the camera recording for \ac{PIV} measurements.

A set of three open-loop actuation mechanisms was tested. The cases were restricted to constant symmetric actuation, with the upstream cylinder kept stationary ($b_1=0$) and the two downstream cylinders rotating with equal and opposite speeds (i.e., $b_2=-b_3$). Under these conditions, the control input can be expressed through the actuation (boat-tailing) parameter:

\begin{equation}
    p=\frac{b_3-b_2}{2}.
\end{equation}

The involved mechanisms correspond to (i) the baseline case ($p=0$), (ii) base-bleeding actuation ($p<0$), in which the cylinder rotation opposes the freestream and generates a jet between the two downstream cylinders, and (iii) boat-tailing actuation ($p>0$), which weakens the gap flow and accelerates the outer shear layer. The parameter $p$ was varied within $-2.8\leq p\leq 2.6$, in steps of $0.2$.

\subsection{Velocity measurements}

\label{subsec:velocity_estimation}
Velocity measurements were carried out using time-resolved \ac{PIV}. The water was seeded with two types of VESTOSINT 2159 polyamide particles with mean diameters of $\SI{56}{\micro\metre}$ and $\SI{11}{\micro\metre}$. The presence of two particle sizes is a residual of previous experimental campaigns, and their complete removal from the closed-loop water tunnel is impractical. Nonetheless, both particle diameters are suitable flow tracers, since they have a Stokes number ($\mathrm{Stk}$) much smaller than unity. The Stokes number is defined as $\mathrm{Stk}=\frac{t_0U_\infty}{2.5D}$, where $t_0=\frac{\rho_pd_p^2}{18\mu}$ is the particle relaxation time, $\rho_p=\SI{1020}{\kilo\gram\per\metre^3}$ is the particle density, $d_p$ is the particle diameter and $\mu=\SI{1}{\milli\pascal\second}$ is the dynamic viscosity of water. In the present experiment, $\mathrm{Stk}=7.3\times10^{-4}$ for the $\SI{56}{\micro\metre}$ particles, and $\mathrm{Stk}=2.8\times10^{-5}$ for the $\SI{11}{\micro\metre}$ particles, ensuring faithful tracking of the flow. The particles were illuminated with a $\SI{450}{\nano\metre}$ LT-40W-AA laser, shaped into a $\SI{1}{\milli\metre}$ horizontal sheet using concave and convex cylindrical lenses. The laser sheet is located at $10D$ from the bottom of the tunnel, far away from the effects of the bottom wall and the free surface.  Images were captured using a $2160\times\SI{2560}{\px}$ Andor Zyla sCMOS camera. To enable time-resolved acquisition, images were cropped to $560\times1100$. The camera was equipped with a Nikon AF Nikkor $\SI{24}{\milli\metre}$ lens and a  $52-\SI{58}{\milli\metre}$ ultra-wide-angle adapter to image a sufficiently large \ac{FOV}. 
The lens aperture was set to $f/\#=8$.
The resulting \ac{FOV} spanned $8D$ in the streamwise direction and $5.5D$ in the cross-stream direction with a spatial resolution of $\SI{0.29}{\milli\metre\per\px}$. 
For each actuation case, 1800 time-resolved snapshots were acquired at $\SI{120}{\hertz}$. Such acquisition frequency corresponds to a time separation $\Delta t=\SI{8.3}{\milli\second}$ between frames, leading to free-stream displacements of approximately $\SI{2.5}{\milli\metre}$ (i.e., $\SI{8.6}{\px}$).

The velocity fields were computed using higher-order multi-frame Particle Tracking Velocimetry \citep[\acs{PTV}, ][]{cierpka2013higher}, employing three sequential images. 
Prior to processing, background illumination was subtracted from the raw images following the removal procedure described by \citet{mendez2017pod}. 
A \ac{PIV}-based prior estimate was employed to guide particle tracking. The interrogation strategy is an iterative multi-grid/multi-pass \citep{soria1996investigation} image deformation algorithm \citep{scarano2001iterativeimage}.
An initial interrogation window of $\SI{144}{\px}$ with 50\% overlap was used to set the reference field. The window size was refined to $\SI{32}{\px}$ with 50\% overlap for the final vector fields.

To precisely determine the incoming flow velocity $U_\infty$, a separate \ac{PIV} acquisition was performed upstream of the cylinders. Since cylinder blockage affects the tunnel speed at any given pumping power, both the laser sheet and the camera were translated approximately $8D$ upstream, so that the \ac{FOV} contained both the unobstructed freestream flow and the cylinders. The known cylinder diameter served as a reference for spatial resolution.

The uncertainty of the incoming flow velocity, estimated from the spatial standard deviation of the upstream measurement region, was below $0.1\%$ of $U_\infty$.

As shown in Figure \ref{fig:2PIV_setup}(a), a Cartesian coordinate system $(x,y)$ was defined in the measurement plane. The origin is located on the symmetry plane of the test section, downstream of the two rear cylinders, at the upstream boundary of the effective \ac{FOV}. This offset accounts for the loss of near-cylinder vectors due to the finite size of the interrogation window. The $x$-axis is aligned with the freestream direction, pointing downstream, and the $y$-axis is oriented perpendicular to the flow in the transverse direction.


\subsection{Force measurements}
\label{subsec:drag_estimation}
Force measurements were performed using three VETEK TCA $\SI{1}{\kilo\gram}$ S-type load cells, capable of measuring static and dynamic loads in tension and compression. This model was selected because the expected maximum drag force remains below its full-scale capacity of $\SI{1}{\kilo\gram}$ and, at the same time, is large enough to provide a reliable electrical output given the rated sensitivity of $\SI{2}{\milli\volt/\volt}$. The analog signals were digitized using a 24-bit ADS1256 analog-to-digital converter. The load cells were mounted on the lower \ac{PVC} plate above the test section (Figure \ref{fig:2PIV_setup}(c)), each positioned with a lateral offset from the center. This arrangement enabled the decoupling of the two in-plane forces, $F_x$ (aligned with the incoming flow) and $F_y$, and the out-of-plane moment $M_z$. 

The load cells produced analog voltage signals that were digitized and converted into forces using a calibration procedure. The calibration accounted for both the applied reference loads and the geometric configuration of the three load cells. Known reference loads were applied to the model, and the resulting voltage responses were used to construct the $3\times3$ calibration matrix $\mathsfbi{K}$, which maps the three load-cell voltages to the two in-plane force components, $F_x$ and $F_y$, and the out-of-plane moment $M_z$. 
Since the three measured quantities have different physical dimensions, the coefficients of the calibration matrix carry the appropriate units, mapping voltages to forces for $F_x$ and $F_y$, and voltages to moments for $M_z$.

Load cell zeroing was performed with the tunnel off at the beginning of each experimental campaign. Following zeroing, the tunnel was switched on, and the load cell output was recorded again for the measurement of the baseline force (without cylinder rotation).
To remove the drift of the force sensor, one should perform load-cell zeroing before each new case.
However, given the slow tunnel startup, it is not possible to stop the tunnel and perform load-cell zeroing before each actuation without possibly introducing a drift in the force measurement. For this reason, the total force was obtained by combining the baseline load and the actuation-induced load. For the actuation-induced load, two measurements were acquired: one with the tunnel running but without actuation, which serves as a substitute for the zero reference; and a second with the prescribed actuation applied. 
This procedure ensures that the baseline load accounts only for the hydrodynamic forces generated by the incoming flow, and the actuation load variation isolates the incremental forces and moment produced by the cylinder rotation.

Once the forces were computed, the dimensionless force coefficient could be obtained. Specifically, the drag coefficient can be expressed as:
\begin{equation}
    C_D =  \frac{F_x}{\frac{1}{2}\rho U_\infty^2 D H},
\end{equation}
where $H$ is the immersed length of each cylinder, equal to $\SI{0.5}{\metre}$.

The uncertainty of the force coefficient arises from several contributions: freestream velocity-measurement uncertainty, load-cell sensor specifications (linearity, hysteresis, and repeatability), calibration uncertainty of the matrix $\mathsfbi{K}$, and statistical uncertainty in the time-averaged load signals. The total uncertainty was obtained through standard error propagation using the effective number of samples calculated from the autocorrelation of the load-cell time series to account for temporal correlation. The complete derivation and the expressions for each uncertainty contribution are provided in \ref{app1}. Since the load cells are highly sensitive to small mechanical and electromagnetic disturbances, including vibrations generated by cylinder rotation, their raw signals contain spurious high-amplitude fluctuations. These fluctuations were not explicitly filtered or removed prior to computing the time-averaged forces; instead, their effect is reflected in the statistical uncertainty of the measurements. As a result, the average relative uncertainty for drag coefficient measurements is approximately $3\%$ of $C_D$.

\subsection{Numerical simulations (\ac{URANS})}
\label{subsec:URANS_methods}

Numerical simulations of the flow around the fluidic pinball were performed using ANSYS Fluent. The geometric dimensions, inter-cylinder spacing, and tunnel wall configuration were matched to the experimental setup. 
In contrast to the experiments, the simulations were performed in two dimensions, on a plane corresponding to that shown in Figure \ref{fig:2PIV_setup}(a). As a result, three-dimensional effects and possible structural vibrations of the cylinders are not captured.

\begin{figure}[!t]
    \centering
    \includegraphics[scale=0.3]{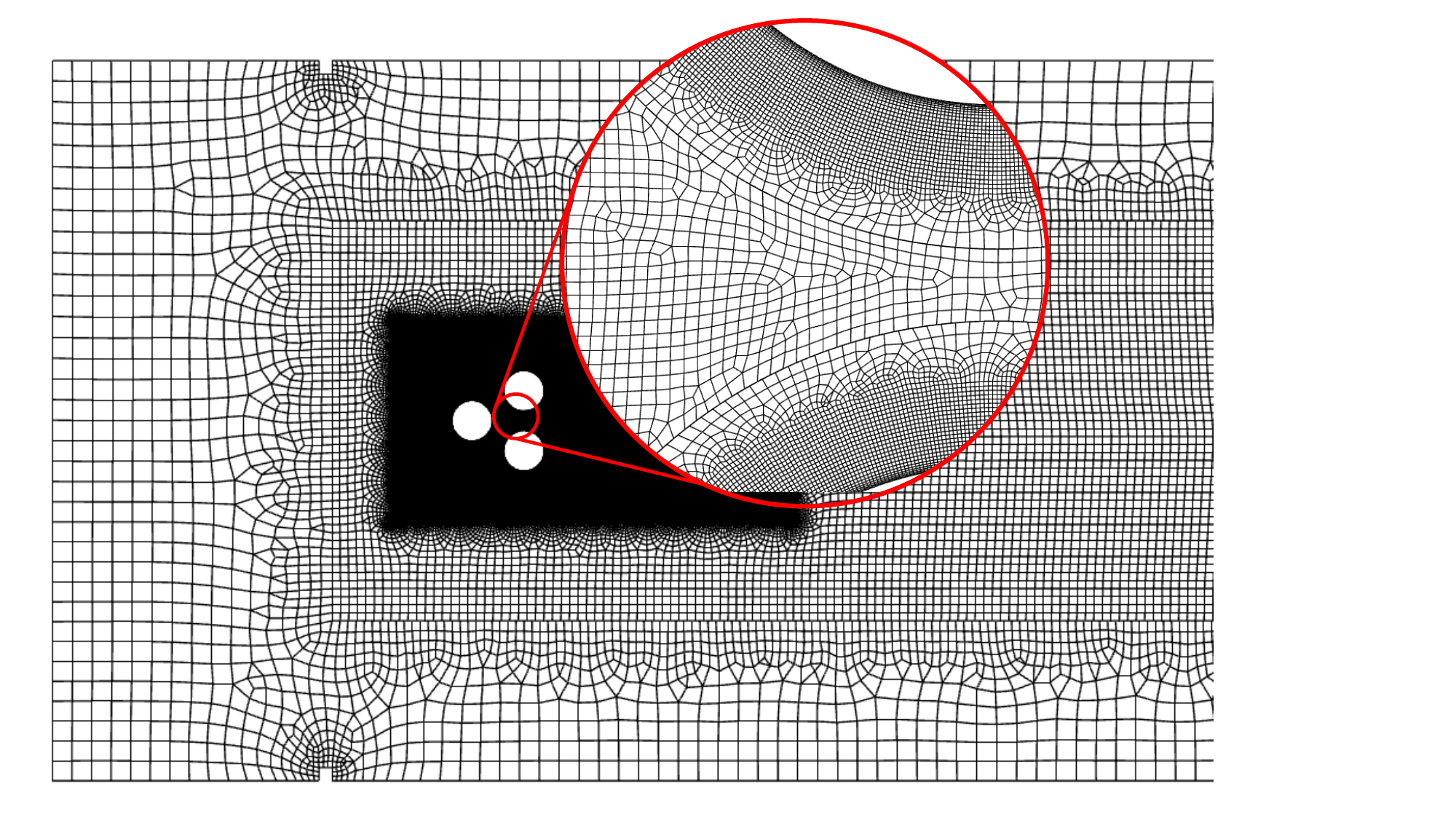}
    \caption{Mesh used in the \ac{URANS} simulations.}
    \label{fig:mesh}
\end{figure}

To ensure adequate spatial resolution, a refined mesh was applied in the vicinity of the cylinder surfaces, as illustrated in Figure \ref{fig:mesh}. The mesh resolution was determined through a grid-convergence study, yielding a final surface spacing of $0.01D$, which provides more than 300 cells along each cylinder circumference. The near-wall resolution was set to $y^+=1$, and 28 inflation layers were applied around each cylinder to resolve the boundary layers without the use of wall functions. This near-wall treatment is optimal for accurately capturing wall-bounded flow behaviour \citep{Benim2008}. The final mesh comprises 180,000 elements and has a minimum orthogonality of 0.58. 

The $k$--$\omega$ SST turbulence model was employed, as it is commonly regarded as one of the most reliable choices for transient RANS simulations involving separated flows \citep{Nishino2008}. Because the configuration involves rotating cylinders, curvature correction was enabled and tuned based on the experimental $p=0$ case. The turbulent viscosity ratio was set to 4, the turbulence intensity to $2\%$, the fluid viscosity to $\nu=\SI{1}{\milli\metre\squared\per\second}$, and the Reynolds number to $Re=9100$ to match the experimental conditions. 

The rotation of the rear cylinders was implemented using local rotating mesh zones, with no-slip boundary conditions prescribed on the rotating walls. The time step for the transient simulations was selected such that the cylinders rotate by one degree per time step. Consequently, the time-step size varies with the actuation level. For these simulations, the parameter $p$ was varied between $-$3 to 2.5, in steps of 0.5.

%% file: Figures/Flow_config.tex
\begin{tikzpicture}[
    >={Stealth[length=2mm]},
    dim/.style={<->, >=latex, thin, color=black},
    axis/.style={->, >=latex, blue, thick},
    arrow_out/.style={->, >=latex, thin, color=black}, 
    x=0.5cm, y=0.5cm 
    ]

    \def\D{1} 
    \def\xshift{-5*\D}
    \def\xshiftb{-4*\D}
    \def\Dc{0.6}
    \def\xshiftc{-2/3*\Dc}

    
    \coordinate (Center1) at (\xshift,0);
    \coordinate (Center2) at (1.2*\D+\xshift, 0.75*\D);
    \coordinate (Center3) at (1.2*\D+\xshift, -0.75*\D);
    
    \draw[thick] (-10*\D, 9*\D) -- (10*\D, 9*\D);
    \draw[thick] (-10*\D, -9*\D) -- (10*\D, -9*\D);
    \fill[gray!30] (-10*\D, 9*\D) rectangle (10*\D, 9.5*\D);
    \fill[gray!30] (-10*\D, -9*\D) rectangle (10*\D, -9.5*\D);

    \draw[fill=orange!40] (-3*\D+\xshift, 9*\D) rectangle ++(-0.3*\D, -0.3*\D);
    \draw[fill=orange!40] (-3*\D+\xshift, -9*\D) rectangle ++(-0.3*\D, 0.3*\D);
    
    \fill[blue!10, opacity=0.8] (2*\D+\xshift, -2.75*\D) rectangle ++(8*\D, 5.5*\D);
    \draw[blue, dashed] (2*\D+\xshift, -2.75*\D) rectangle ++(8*\D, 5.5*\D);
    \node[font=\footnotesize,blue, below] at (6*\D+\xshift, -2.75*\D) {\small PIV Field Of View};

    \foreach \c in {Center1, Center2, Center3} {
        \draw[fill=green!20, thick] (\c) circle (0.5*\D); 
        \draw[dash dot, thin] (\c) -- ++(-0.8*\D,0) -- ++(1.6*\D,0); 
        \draw[dash dot, thin] (\c) -- ++(0,-0.8*\D) -- ++(0,1.6*\D); 
    }


    \draw[thick, green!40!black](Center1) ++(120:0.7*\D) arc (120:240:0.7*\D)node[font=\footnotesize,near end, left] {$b_1$};
    \path (Center1) ++(120:0.7*\D) coordinate (start) arc (120:240:0.7*\D) coordinate (end);
    \draw[-{Stealth[scale=0.75]}, line width=0.8pt,green!40!black]
    (end) -- ++(-40:0.2);

    \draw[thick, green!40!black](Center2) ++(30:0.7*\D) arc (30:150:0.7*\D)node[font=\footnotesize,midway, above] {$b_2$};
    \path (Center2) ++(30:0.7*\D) coordinate (start) arc (30:150:0.7*\D) coordinate (end);
    \draw[-{Stealth[scale=0.75]}, line width=0.8pt,green!40!black]
    (end) -- ++(-130:0.2);

    \draw[thick, green!40!black](Center3) ++(-150:0.7*\D) arc (-150:-30:0.7*\D)node[font=\footnotesize,midway, below] {$b_3$};
    \path (Center3) ++(-150:0.7*\D) coordinate (start) arc (-150:-30:0.7*\D) coordinate (end);
    \draw[-{Stealth[scale=0.75]}, line width=0.8pt,green!40!black]
    (end) -- ++(50:0.2);

    \draw[dim] (9*\D, 9*\D) -- (9*\D, -9*\D) node[font=\footnotesize,near end, above, rotate=90] {$18D$};
    \draw[dim] (2*\D+\xshift, 3.5*\D) -- (8*\D+\xshift+2*\D, 3.5*\D) node[font=\footnotesize,midway, above] {$8D$};
    \draw[blue, thin] (2*\D+\xshift, 2.8*\D) -- (2*\D+\xshift, 3.7*\D);
    \draw[blue, thin] (8*\D+\xshift+2*\D, 2.8*\D) -- (8*\D+\xshift+2*\D, 3.7*\D);

    \draw[thin] (Center2) -- (3.7*\D+\xshift, 0.75*\D);
    \draw[thin] (Center3) -- (3.7*\D+\xshift, -0.75*\D);
    \draw[thin] (3.5*\D+\xshift, 0.75*\D) -- (3.5*\D+\xshift, -0.75*\D);
    \draw[arrow_out] (3.5*\D+\xshift, -1.25*\D) -- (3.5*\D+\xshift, -0.75*\D); 
    \draw[arrow_out] (3.5*\D+\xshift, 2.5*\D) -- (3.5*\D+\xshift, 0.75*\D); 
    \node[font=\footnotesize,rotate=90] at (3*\D+\xshift, 1.75*\D){$1.5D$};

    \draw[dim] (9*\D+\xshift+2*\D, -2.75*\D) -- (9*\D+\xshift+2*\D, 2.75*\D) node[font=\footnotesize,near end, above, rotate=90] {$5.5D$};
    \draw[blue, thin] (8*\D+\xshift+2*\D, 2.75*\D) -- (9.2*\D+\xshift+2*\D, 2.75*\D);
    \draw[blue, thin] (8*\D+\xshift+2*\D, -2.75*\D) -- (9.2*\D+\xshift+2*\D, -2.75*\D);
    \draw[dim] (-3*\D+\xshift, 10*\D) -- (0+\xshift, 10*\D) node[font=\footnotesize,midway, above] {$3D$};
    \draw[thin] (-3*\D+\xshift, 9*\D) -- (-3*\D+\xshift, 10.5*\D);
    \draw[thin] (0+\xshift, 0) -- (0+\xshift, 10.5*\D);

    
    \draw[thin] (-3.3*\D+\xshift, -9*\D) -- (-4.5*\D+\xshift, -9*\D);   
    \draw[thin] (-3.3*\D+\xshift, -8.7*\D) -- (-4.5*\D+\xshift, -8.7*\D);
    
    \draw[thin] (-4.2*\D+\xshift, -9*\D) -- (-4.2*\D+\xshift, -8.7*\D);
    
    \draw[arrow_out] (-4.2*\D+\xshift, -9.5*\D) -- (-4.2*\D+\xshift, -9*\D); 
    \draw[arrow_out] (-4.2*\D+\xshift, -7.2*\D) -- (-4.2*\D+\xshift, -8.7*\D); 
    
    \node[font=\footnotesize,rotate=90] at (-4.6*\D+\xshift, -7.8*\D) {$0.3D$};

    \draw[thin] (-3.3*\D+\xshift, -9*\D) -- (-3.3*\D+\xshift, -7.8*\D); 
    \draw[thin] (-3.0*\D+\xshift, -9*\D) -- (-3.0*\D+\xshift, -7.8*\D); 
    
    \draw[thin] (-3.3*\D+\xshift, -8.2*\D) -- (-3.0*\D+\xshift, -8.2*\D);
    
    \draw[arrow_out] (-3.8*\D+\xshift, -8.2*\D) -- (-3.3*\D+\xshift, -8.2*\D); 
    \draw[arrow_out] (-1.5*\D+\xshift, -8.2*\D) -- (-3.0*\D+\xshift, -8.2*\D); 
    
    \node [font=\footnotesize] at (-3.15*\D+0.65*\xshift, -7.8*\D) {$0.3D$};

    \draw[dashed] (-10*\D, 0) -- (10*\D, 0);

    \draw[->, thick] (-9*\D, 2*\D) -- (-6*\D, 2*\D) node[font=\footnotesize,midway, above] {$U_\infty$};
    \draw[axis] (2*\D+\xshift, 0) -- (3*\D+\xshift, 0) node[font=\footnotesize,above] {$x$};
    \draw[axis] (2*\D+\xshift, 0) -- (2*\D+\xshift, 1*\D) node[font=\footnotesize,right] {$y$};
    \draw[blue, fill=white] (2*\D+\xshift, 0) circle (0.2*\D);
    \draw[blue, fill=blue] (2*\D+\xshift, 0) circle (0.075*\D)node[font=\footnotesize,below right,blue] {$z$};

    \node[font=\bfseries, anchor=south west] at (-10*\D, 10*\D) {(a)};

    \begin{scope}[xshift=10cm, yshift=-0.75cm] 
        
        \draw[thick] (-9.5*\D, 0.5*\D) -- (5*\D, 0.5*\D);
        \draw[thick] (-9.5*\D, -10*\D) -- (5*\D, -10*\D);
        \fill[gray!30] (-9.5*\D, 0.5*\D) rectangle (5*\D, 0.75*\D);
        \fill[gray!30] (-9.5*\D, -10*\D) rectangle (5*\D, -10.25*\D);

        \draw[fill=orange!40, draw=brown] (-4*\D+\xshiftb/2, 0.5*\D) rectangle (-3.8*\D+\xshiftb/2, -10*\D);
        \node[align=right, font=\footnotesize] at (-6*\D+\xshiftb/2, 0) {Turbulators};
        \draw[->, thin, brown] (-4.3*\D+\xshiftb, -0.25) -- (-4.05*\D+\xshiftb/2, -1.25*\D);

        \draw[line width=2.5pt] (-1.8*\D+\xshiftb, 1.5*\D) -- (-1.8*\D+\xshiftb, 13*\D);
        \draw[line width=2.5pt] (1.8*\D+\xshiftb, 1.5*\D) -- (1.8*\D+\xshiftb, 13*\D);
        \draw[thick] (-2.2*\D+\xshiftb, 11.5*\D) -- (2.2*\D+\xshiftb, 11.5*\D);

        \draw[fill=blue!20, draw=black] (-0.1*\D+\xshiftb, 11.5*\D) rectangle (-0.7*\D+\xshiftb, 12.5*\D);
        \draw[fill=blue!20, draw=black] (0.1*\D+\xshiftb, 11.5*\D) rectangle (0.7*\D+\xshiftb, 12.5*\D);
        
        \node[font=\footnotesize, anchor=east] at (5*\D+\xshiftb, 13.5*\D) {Stepper motors};
        \draw[->, blue!50!black] (1.55*\D+\xshiftb, 13.3*\D) -- (0.55*\D+\xshiftb, 12.5*\D);

        \draw[fill=green!20, draw=green!20!black] (-0.65*\D+\xshiftb, 0.5*\D) rectangle (-0.15*\D+\xshiftb, -9.8*\D);
        \draw[fill=green!20, draw=green!20!black] (0.15*\D+\xshiftb, 0.5*\D) rectangle (0.65*\D+\xshiftb, -9.8*\D);

        \fill[black] (-0.5*\D+\xshiftb, 11.5*\D) rectangle (-0.3*\D+\xshiftb, -9.5*\D);
        \fill[black] (0.3*\D+\xshiftb, 11.5*\D) rectangle (0.5*\D+\xshiftb, -9.5*\D);

        \filldraw[fill=red!40, draw=red!80!black] (-1.3*\D+\xshiftb, 3.2*\D) rectangle (-0.9*\D+\xshiftb, 2.1*\D); 
        
        \draw[fill=gray!20, thick] (-2*\D+\xshiftb, 7.5*\D) rectangle (2*\D+\xshiftb, 8.2*\D);
        \draw[fill=gray!20, thick] (-2*\D+\xshiftb, 1.5*\D) rectangle (2*\D+\xshiftb, 2.2*\D);
        
        \filldraw[fill=red!40, draw=red!80!black] (1.1*\D+\xshiftb, 3.2*\D) rectangle (1.5*\D+\xshiftb, 2.1*\D);    
        
        \draw[fill=red!40, draw=red!80!black] (2*\D+\xshiftb, 2.1*\D) rectangle (3*\D+\xshiftb, 3.2*\D);
        \draw[red!80!black] (2.3*\D+\xshiftb, 2.1*\D) -- (2.3*\D+\xshiftb, 2.6*\D); 
        \draw[red!80!black] (2.7*\D+\xshiftb, 2.6*\D) -- (2.7*\D+\xshiftb, 3.2*\D); 
        
        \node[right, font=\footnotesize] at (3.5*\D+\xshiftb, 2.4*\D) {Load cells};
        \draw[->, red!80!black] (3.75*\D+\xshiftb, 2.3*\D) -- (3*\D+\xshiftb, 2.7*\D);
         \draw[->, red!80!black] (6.4*\D+\xshiftb, 2.3*\D) -- (7*\D+\xshiftb, 3.2*\D);

        \node[right, font=\footnotesize] at (1.5*\D+\xshiftb, -1.5*\D) {Cluster of cylinders};
        \draw[->, green!40!black] (1.5*\D+\xshiftb, -1.5*\D) -- (0.8*\D+\xshiftb, -2*\D);

        \draw[blue, thick] (0.5*\D+\xshiftb, -5*\D) -- (6.5*\D+\xshiftb, -5*\D);
        \node[below, font=\footnotesize] at (5.5*\D+\xshiftb, -6*\D) {Laser sheet};
        \draw[->, blue] (5.5*\D+\xshiftb, -6*\D) -- (4*\D+\xshiftb, -5.2*\D);

        \draw[dim] (8.75*\D+\xshiftb, 0.5*\D) -- (8.75*\D+\xshiftb, -10*\D) node[font=\footnotesize, midway, above, rotate=90] {$17D$};
        \draw[dim] (3*\D+\xshiftb,-5*\D) -- (3*\D+\xshiftb, -10*\D) node[font=\footnotesize, midway, above, rotate=90] {$10D$};

        \draw[->,thick] (-9*\D, -5*\D) -- (-6.5*\D, -5*\D);
        \node[font=\footnotesize,above] at (-8*\D, -5*\D) {$U_\infty$};

        \node[font=\bfseries, anchor=south west] at (-9.5*\D, 13*\D) {(b)};

    \end{scope}

    \begin{scope}[xshift=11cm, yshift=2.5cm]

        \draw[fill=gray!20, thick] (-2*\D, -2*\D) rectangle (2*\D, 2*\D);

        \draw[fill=black, thick] (-1.75*\D,-1.75*\D) circle (0.1*\D); 
        \draw[fill=black, thick] (-1.75*\D,1.75*\D) circle (0.1*\D);
        \draw[fill=black, thick] (1.75*\D,-1.75*\D) circle (0.1*\D);
        \draw[fill=black, thick] (1.75*\D,1.75*\D) circle (0.1*\D);

        \filldraw[fill=red!40, draw=red!80!black] (-1.3*\D, 3.1*\D) rectangle (-0.9*\D, 2*\D); 
        \draw[red!80!black] (-1.3*\D, 2.8*\D) -- (-0.9*\D, 2.8*\D); 
        \draw[dashed] (-1.1*\D,3.3*\D) -- (-1.1*\D,1.8*\D);

        \filldraw[fill=red!40, draw=red!80!black] (0.9*\D, -2*\D) rectangle (1.3*\D, -3.1*\D);    
        \draw[red!80!black] (1.3*\D, -2.8*\D) -- (0.9*\D, -2.8*\D);
        \draw[dashed] (1.1*\D,-3.3*\D) -- (1.1*\D,-1.8*\D); 
        
        \draw[fill=red!40, draw=red!80!black] (2*\D, 1.3*\D) rectangle (3*\D, 0.9*\D);
        \draw[red!80!black] (2.7*\D, 1.3*\D) -- (2.7*\D, 0.9*\D);
        \draw[dashed] (1.8*\D,1.1*\D) -- (3.2*\D,1.1*\D); 

        \coordinate (Center1) at (\xshiftc,0);
        \coordinate (Center2) at (1.2*\Dc+\xshiftc, 0.75*\Dc);
        \coordinate (Center3) at (1.2*\Dc+\xshiftc, -0.75*\Dc);

        \foreach \c in {Center1, Center2, Center3} {
        \draw[fill=green!20, thick] (\c) circle (0.5*\Dc); 
        \draw[fill=black, thick] (\c) circle (0.1*\D); 
        \draw[dash dot, thin] (\c) -- ++(-0.8*\Dc,0) -- ++(1.6*\Dc,0); 
        \draw[dash dot, thin] (\c) -- ++(0,-0.8*\Dc) -- ++(0,1.6*\Dc); 
        }   

        \draw[dim] (-1.1*\D, 4*\D) -- (-2, 4*\D) node[font=\footnotesize,midway, above] {$1.15D$};
        \draw[thin] (-1.1*\D, 3.1*\D) -- (-1.1*\D, 4*\D);
        \draw[thin] (-2, 0) -- (-2, 4*\D);

        \node[font=\bfseries, anchor=south west] at (-3.5*\D, 2.5*\D) {(c)};

    \end{scope}

\end{tikzpicture}

%% file: Chapters/3-Results.tex
\section{Results} \label{sec:results}

This section presents the results of both the experimental campaign and the numerical simulations. The time-averaged flow fields and a modal analysis that characterizes the wake dynamics are presented first. The analysis of the inter-cylinder region from the numerical simulations is then outlined. Finally, the resulting drag coefficient over the investigated range of actuation parameters is examined.

\subsection{Wake flow development and time-averaged fields} \label{subsec:wakeflow}

\begin{figure*}[!t]
    \centering
    \input{Figures/FP_WakeFlowDevelopment.tex}
    \caption{Contour representation of the time-averaged velocity magnitude in the wake of the fluidic pinball under symmetric actuation, with superposed \ac{LIC} representation of the mean velocity field. Dashed colored lines identify the actuation cases. For the unforced case (a,d), the gap jet deflects upward or downward, producing two quasi-steady asymmetric wakes. Increasing $p$ (b,c) suppresses the gap jet and narrows the wake, whereas decreasing $p$ (e,f) strengthens a central jet that divides the wake into two lateral branches.}
    \label{fig:4wakeflow_res}
\end{figure*}
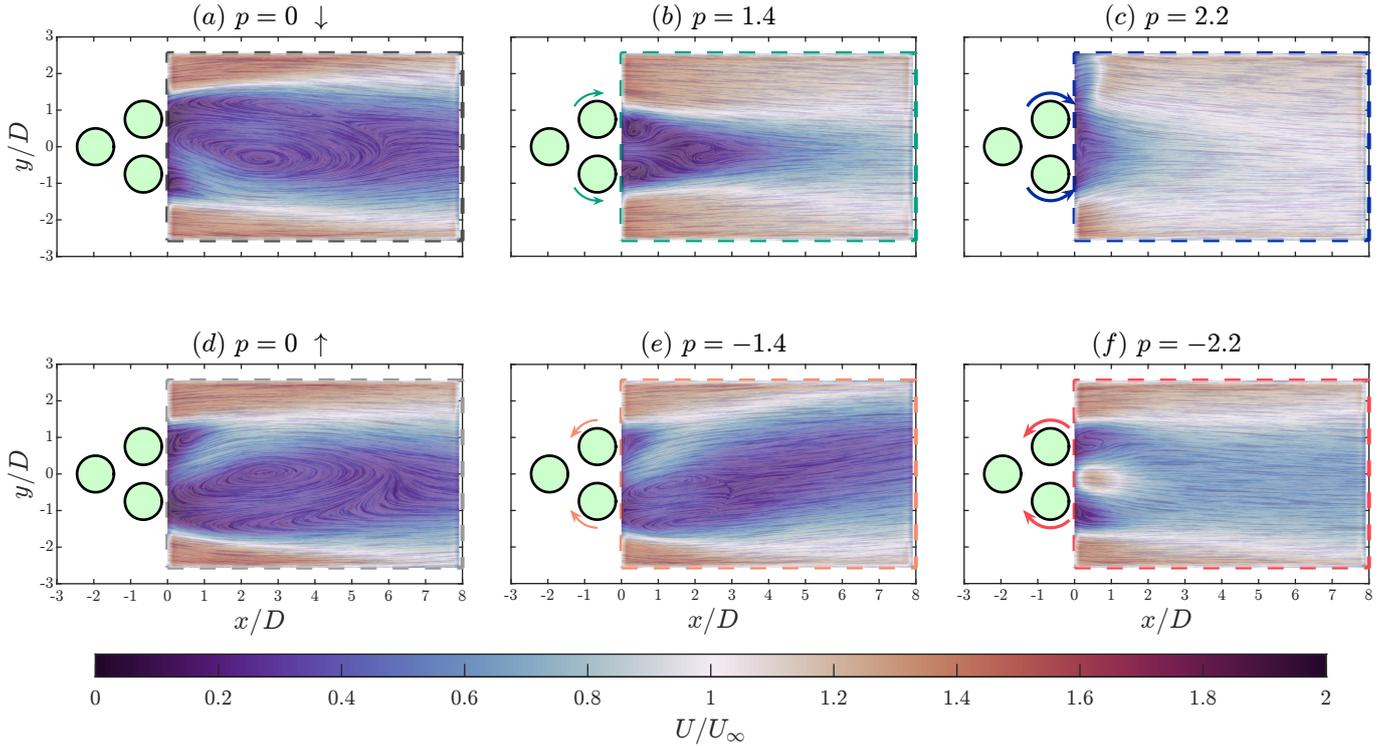

Figure \ref{fig:4wakeflow_res} shows representative time-averaged velocity fields for several actuation levels, superposed with their corresponding Line Integral Convolution \citep[\acs{LIC}, ][]{forssell1995using} representation. Each subfigure illustrates the characteristic wake topology obtained for baseline, boat-tailing, and base-bleeding actuation regimes.

Figure~\ref{fig:4wakeflow_res}(a) and (d) display the two flow topologies for the unforced flow.
They both describe an asymmetric flow with a deflected jet, upwards ($\uparrow$) for Figure~\ref{fig:4wakeflow_res}(a) and downwards ($\downarrow$) for Figure~\ref{fig:4wakeflow_res}(d).
The gap jet emerging between the rear cylinders deflects either upward or downward, producing two mirror-symmetric but stable wake topologies. Both states are dominated by a large recirculation region downstream of the cluster of cylinders. During the experiments, no sudden switching was observed; however, it was verified that a short transient with a sufficiently strong actuation input (for instance, a sudden rotation of a cylinder) could force the wake to settle into either branch, confirming the sensitivity of the system to initial conditions and external inputs.

Under boat-tailing actuation ($p>0$), the rotation of the downstream cylinders accelerates the outer shear layers and suppresses the gap flow, enabling wake symmetrization. For actuation with moderate intensity (Figure \ref{fig:4wakeflow_res} (b), $p=1.4$), the separated region is substantially reduced, and the wake becomes narrow and centered. As the actuation strength further increases (Figure \ref{fig:4wakeflow_res} (c), $p=2.2$), the wake elongates, the outer flow reattaches more slowly, and the incoming velocity recovers only beyond $x\approx4D$.

Conversely, base-bleeding actuation ($p<0$) enhances the jet between the rear cylinders. For relatively mild actuation (Figure \ref{fig:4wakeflow_res} (e), $p=-1.4$), the wake still resembles the asymmetric baseline state, with a wide recirculation region and a weak inclined gap jet. At stronger actuation levels (Figure \ref{fig:4wakeflow_res} (f), $p=-2.2$), the jet becomes well-defined and further centered between the rear cylinders, resulting in a markedly altered wake structure with reduced recirculation and a clear outward deflection of the shear layers.

Overall, the inspection of the time-averaged flow fields reveals that both positive and negative $p$ progressively suppress the asymmetric bistable regime. While $p>0$ contracts the wake, $p<0$ enhances the gap jet and widens the wake. These observations confirm that symmetric rotation has a clear control authority over the wake structure in the turbulent regime.

\subsection{Modal analysis} \label{subsec:POD}

In the previous section, the wake was shown to exhibit strong recirculation and unsteady vortex shedding, whose intensity varies across actuation regimes. To quantify these changes more systematically, \ac{POD} was performed on the time-resolved velocity fields \citep{Holmes2012book}.  
The decomposition was performed on the fluctuating component of the flow, obtained by subtracting the time-averaged velocity fields from each set of instantaneous measurements.

\begin{figure}[!t]
    \centering
    \includegraphics{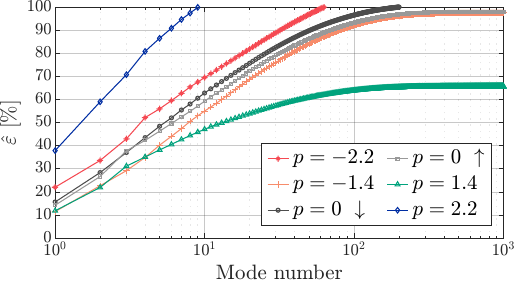}
    \caption{Cumulative normalized \ac{POD} energy as a function of mode number for six actuation cases: base bleeding, unforced, and boat tailing. The energy of the first 200 modes of the baseline case $p=0\ \downarrow$ is used for the normalization.}
    \label{fig:4POD_cumsum}
\end{figure}

To enable a direct comparison of the energetic content of each actuated case, the cumulative normalized \ac{POD} energy is defined as:
\begin{equation}
    \hat{\varepsilon}(m) = \frac{\sum^{m}_{k=1}\lambda_k}{\sum^{N_{\mathrm{ref}}}_{k=1}\lambda_k^{(0)}},
    \label{eq:lambda_norm}
\end{equation}
where $\lambda_k$ are the \ac{POD} eigenvalues, and the denominator corresponds to the total energy contained in the first $N_{\mathrm{ref}}=200$ modes of the baseline case $p=0\ \downarrow$ (asymmetric downward gap jet), which corresponds to $99\%$ of its energy. With this normalization, equation \eqref{eq:lambda_norm} provides a direct comparison of the energetic content of each actuated flow relative to the non-actuated configuration. Figure \ref{fig:4POD_cumsum} presents this parameter for each actuated case.

For the baseline flow ($p=0$), the gap-upward and gap-downward configurations exhibit nearly identical energy distribution, confirming the statistical equivalence of the two quasi-stable wake states. The minor differences between them are probably due to the non-perfectly symmetric spanwise imaging of the \ac{PIV} plane. In either case, approximately 90\% of the energy is recovered within the first 70 modes, reflecting the high level of unsteadiness associated with vortex shedding. 

Base-bleeding cases ($p=-1.4$ and $p=-2.2$) show a similarly gradual rise of the cumulative energy curve. In particular, the strong base-bleeding case exhibits larger total energy content, indicating that the enhanced gap jet introduces additional energetic fluctuations and sustains unsteady wake dynamics.

On the other hand, weak boat-tailing actuation ($p=1.4$) results in a lower cumulative energy relative to the baseline. This reflects wake stabilization and vortex shedding attenuation, leading to a less energetic flow, i.e., with a lower turbulence kinetic energy. Nonetheless, for strong boat-tailing actuation ($p=2.2$), the cumulative energy exceeds that of the baseline case. Although the mean wake appears stabilized, this indicates the emergence of energetic unsteady structures.

\begin{figure*}[!t]
    \centering
    \input{Figures/FP_POD_phi_all.tex}
    \caption{First \ac{POD} spatial mode and leading oscillatory pair associated with the dominant vortex-shedding dynamics, together with their corresponding temporal coefficients for representative actuation cases: (a) base bleeding $p=-2.2$, modes 1-3; (b) baseline $p=0$, modes 1-3; (c) weak boat tailing $p=1.4$, modes 1-2; (d) strong boat tailing $p=2.2$, modes 1-2. Line styles and colors uniquely identify the \ac{POD} modes. The same color and line style are consistently used for each spatial mode and its corresponding temporal coefficient.}
    \label{fig:4POD_phi}
\end{figure*}
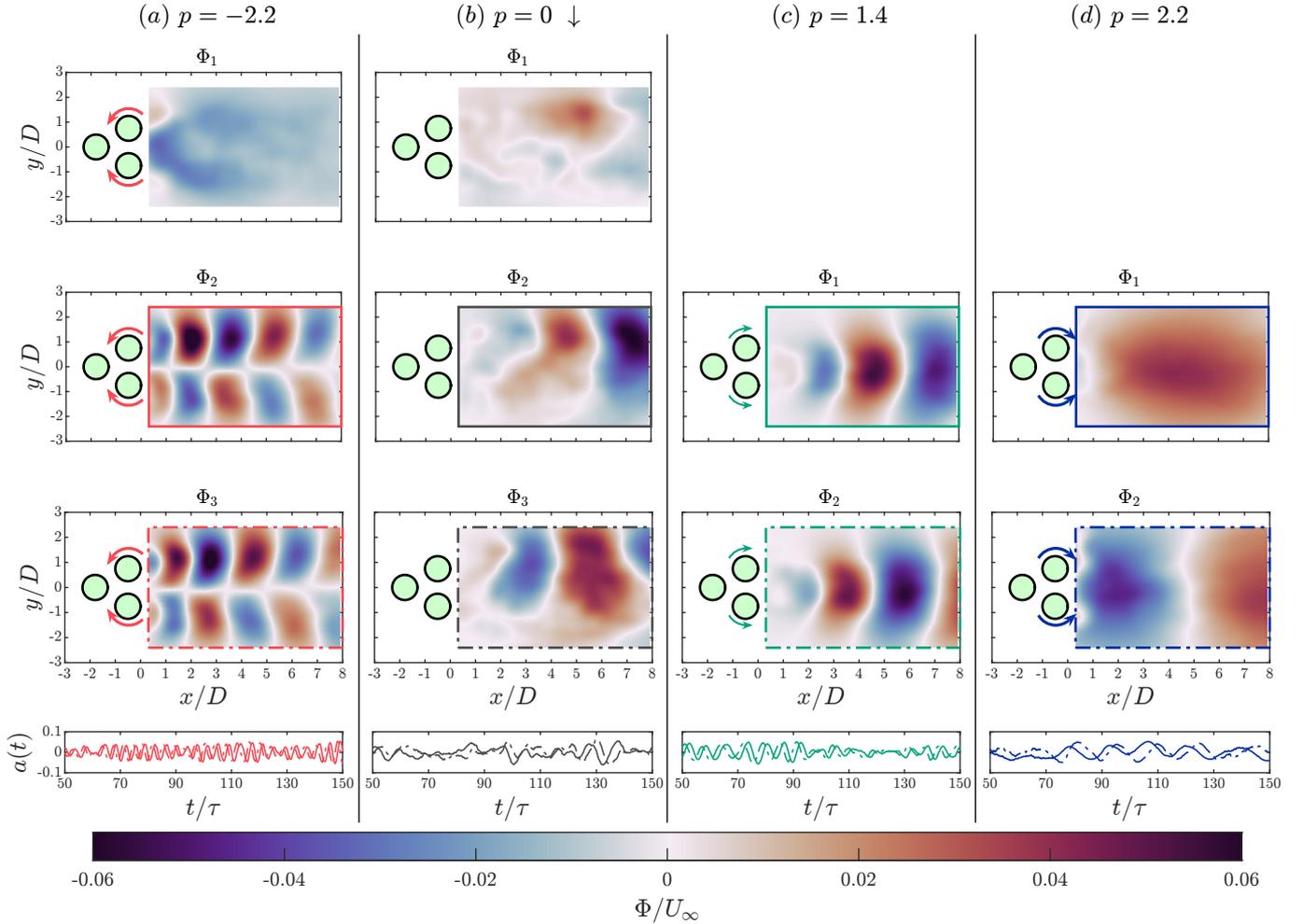

Figure \ref{fig:4POD_phi} shows representative \ac{POD} spatial modes for the three actuation mechanisms. For each case, the modes displayed include the leading \ac{POD} mode and the first oscillatory mode pair associated with vortex shedding. 

Due to the appearance of the dominant oscillatory dynamics at different positions in the modal hierarchy for different actuation cases, the oscillatory pair corresponds to modes 2 and 3 for the base-bleeding (a) and baseline (b) cases, and to modes 1 and 2 for the weak (c) and strong (d) boat-tailing cases. The oscillatory modes were identified based on the spectral content of their temporal coefficients. The spatial modes are arranged to facilitate comparison across actuation regimes, and the corresponding temporal coefficients are shown in the bottom row to highlight the periodic dynamics of the vortex-shedding.

For the base-bleeding (a) and baseline (b) cases, their respective first \ac{POD} mode represents lateral shifts of the gap jet, while the periodic vortex shedding is captured by the second and third modes. Although the mean flow was removed, the gap jet remains deflected toward one side and undergoes slow fluctuations about that state. These variations are not captured by the time-averaged field and therefore emerge as the leading \ac{POD} mode. 

On the contrary, the symmetric wake and the dominant vortex-shedding dynamics of boat-tailing actuation (c, d) result in the emergence of the oscillatory mode pair as modes 1 and 2. 

For the base-bleeding case (a) $p=-2.2$, modes 2 and 3 reveal a clear splitting of the vortex-shedding structures on either side of the gap jet. The corresponding temporal coefficients display a well-defined periodic oscillation, indicating persistent unsteady dynamics. The wake splitting reduces the length scale of the dominant flow structures with respect to the baseline case, leading to shorter characteristic time scales and, therefore, higher dominant frequencies. Additionally, the gap jet sustains strong shear-layer instabilities, explaining the increased energetic content regarding the baseline regime observed in Figure \ref{fig:4POD_cumsum}.

The baseline case (b) $p=0$ exhibits the classical anti-symmetric vortex-shedding pattern. Modes 2 and 3 form a conjugate oscillatory pair, as confirmed by their phase-shifted temporal coefficients. The spatial symmetry of the modes is partially degraded, especially in the near-wake region, due to the deflection of the gap jet.

\begin{figure*}[ht]
    \centering
    \input{Figures/URANS_vorticity}
    \caption{Snapshots of the URANS results showing the vorticity field ($\omega$) around the fluidic pinball under symmetric actuation. As the boat-tailing actuation increases, the front cylinder wake moves outwards, the separation point on the rear cylinders moves downstream, and the wake width reduces until $p=1.5$. For $p>1.5$, the wake pattern changes significantly, with only one cylinder shedding a vortex at a time.}
    \label{fig:URANS_vorticity}
\end{figure*}
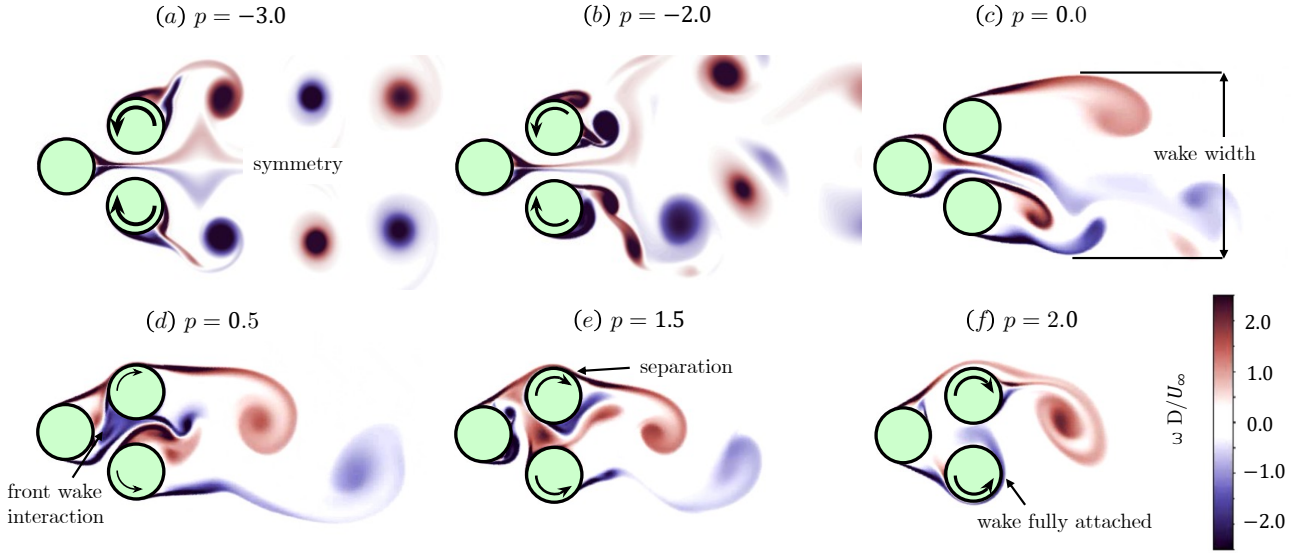

For boat-tailing actuation, both the weak (c) $p=1.4$ and strong (d) $p=2.2$ show that the mean recirculation bubble is significantly shortened and narrowed. Nonetheless, the \ac{POD} analysis reveals that coherent vortex-shedding structures persist and are dominant in the first two \ac{POD} modes. This indicates that, although boat tailing strongly reduces the amplitude of unsteady fluctuations and leads to drag reduction, it does not completely suppress time-dependent wake dynamics. 

A clear distinction between the two boat-tailing regimes appears in the temporal coefficients. For $p=1.4$, the shedding frequency stays comparable to that of the baseline flow, indicating the interaction of the three cylinders. In contrast, for $p=2.2$, the dominant frequency is substantially lower, approximately an order of magnitude smaller than in the baseline. This shift to lower-frequency dynamics confirms that the fluidic pinball behaves increasingly as a single bluff body rather than as a set of interacting cylinders.

In summary, the \ac{POD} analysis demonstrates that the actuation mechanisms substantially modify the energetic distribution and dominant flow structures of the wake. While boat-tailing actuation significantly reduces wake complexity and drag, periodic dynamics remain detectable in the leading \ac{POD} modes. This highlights the distinction between wake stabilization in a mean-flow sense and the complete suppression of unsteady vortex shedding.

\subsection{Near-cylinder flow inspection} \label{subsec:URANS}

The experimental measurements enable a detailed characterization of the wake. However, the flow in the immediate vicinity of the cylinder surfaces cannot be captured due to inherent limitations of the \ac{PIV} technique, such as laser reflections. To provide a more complete picture of the flow physics governing the interaction with the fluidic pinball, \ac{URANS} simulations were performed for the same configurations investigated experimentally. These results are interpreted with caution, as \ac{URANS} is known to have limitations in separated flows \citep{Nishino2008}. For this reason, the viscous and turbulence models were carefully selected, as described in Section \ref{subsec:URANS_methods}, and only the major trends across actuation regimes are considered physically meaningful.

Figure \ref{fig:URANS_vorticity} shows instantaneous vorticity fields extracted from the transient simulations. The snapshots correspond to a fully developed wake and were selected such that the upper cylinder is in the process of shedding a vortex. Consistent with the experimental observations, vortex shedding alternates between the upper and lower rear cylinders for actuated cases, whereas the baseline configuration ($p=0$) exhibits a preferred deflection direction. In the present simulations, the wake preferentially deflects upward, consistent with the $p=0\ \downarrow$ case shown in Figure \ref{fig:4wakeflow_res}. For this non-actuated configuration, separation occurs approximately at the midline of the rear cylinders. In addition, the wake originating from the front cylinder is squeezed between the two rear cylinders, forming a pronounced shear layer in the gap region.

As the actuation parameter increases to $p=0.5$, the wake width decreases. This behavior is associated with a downstream shift of the separation point induced by cylinder rotation and with a corresponding reduction in drag, as shown later in Section \ref{subsec:forces}. These trends are consistent with both the experimental results and the expected flow physics. At this actuation level, the wake of the front cylinder still passes primarily through the gap between the rear cylinders. However, it begins to interact with them, alternating between the two.

As the strength of boat-tailing actuation increases further to $p=1.5$, the wake continues to contract, accompanied by additional drag reduction. However, a portion of the front-cylinder wake begins to bypass the gap and instead convects around the outer sides of the rear cylinders. At higher values of $p$, the wake stops contracting, and the flow topology changes substantially. The entire wake of the front cylinder is diverted around the rear cylinders. At the same time, vortex formation and shedding become slower, leading to a reduced shedding frequency. At a given instant, one rear cylinder is experiencing flow separation, while the flow on the other remains largely attached. This decrease in vortex-shedding frequency is consistent with the trends identified in the \ac{POD} analysis presented in Section \ref{subsec:POD}.

For base-bleeding actuation, the simulations predict a widening of the wake. In the range $-2 < p < 0$, vortex shedding is asymmetric. The separation point remains located at the midline of the rear cylinders, where strong coherent vortices are formed and shed. Meanwhile, the wake of the front cylinder narrows.

At stronger actuation, around $p \approx -3$, the wake of the entire system becomes symmetric. Equal counter-rotating vortices are shed from the rear cylinders, while the front-cylinder wake widens and adopts a rhomboidal shape after passing between them. Notably, the separation point on the rear cylinders shifts downstream, even though the surface rotates in the forward direction.

\subsection{Actuation effects on the drag coefficient} \label{subsec:forces}

\begin{figure}[!t]
    \centering
    \includegraphics{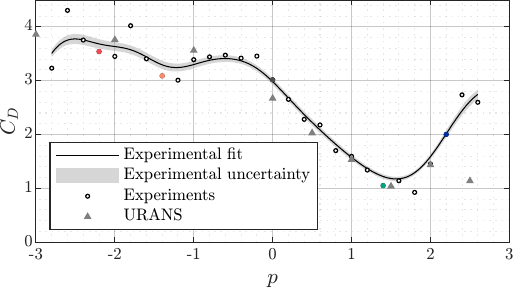}
    \caption{Drag coefficient as a function of the actuation parameter $p$ for baseline, base-bleeding, and boat-tailing actuation cases. Empty circular markers denote experimental measurements. The solid black line corresponds to the local linear regression estimates of the experimental drag data, while the shaded gray envelope represents the measurement uncertainty associated with the force measurements. Selected actuation cases of interest are highlighted using the same color coding as in previous figures. Gray triangle markers indicate the drag coefficients obtained from the \ac{URANS}.}
    \label{fig:4forces}
\end{figure}

Having established the effect of the actuation on the wake structure and its dynamics, we now examine its impact on the aerodynamic force. Figure \ref{fig:4forces} shows the drag coefficient for the analyzed actuation cases, including both experimental measurements and numerical simulations. The experimental data are represented by the median values, providing a robust value despite occasional outliers from measurement noise and load-cell sensitivity. Local linear regression estimates, computed using a Gaussian kernel and a plug-in bandwidth selector \citep{ruppert1995effective}, reveal the underlying trend of the drag. The measurement uncertainty is shown as a shaded band around the fitted curve, obtained by smoothing the measurement uncertainty itself. Cases of interest are highlighted using the same color scheme as in previous figures: the baseline case in dark gray, boat-tailing actuation cases in cool tones, and base-bleeding cases in warm tones. Overall, \ac{URANS} and experimental data show good agreement.


The drag coefficient exhibits a strong and non-monotonic dependence on the actuation parameter. 
Overall, the force trends are consistent with the modal analysis presented in Section \ref{subsec:POD} and the inter-cylinder features shown in Section \ref{subsec:URANS}. 
Base-bleeding actuation ($p<0$) leads to a gradual but consistent increase in drag relative to the baseline, showing the strengthening of the gap jet and the persistence of noticeable recirculation in the wake, as shown in Figure \ref{fig:4wakeflow_res}. At $p=-2.6$, a slight local reduction in the experimental drag is observed, although the value remains above the baseline. This local decrease coincides with wake symmetrization, which stabilizes the flow and partially mitigates drag increase (see Section \ref{sec:ROM}). This local decrease is not captured by the numerical results since the range is more sparse to attain the general trend.

In contrast, boat-tailing actuation ($p>0$) produces a pronounced drag reduction for moderate values of $p$. The drag decreases steadily up to an optimal value around $p\approx1.8$, where the wake is most effectively stabilized and narrowed. This result is consistent with Figure \ref{fig:4wakeflow_res}. Beyond this point, further increases in actuation strength no longer improve performance. Instead, the drag coefficient increases again, although it remains below the baseline value. This behavior indicates a loss of control authority beyond the minimum-drag boat-tailing regime. The drag increase might be further explained with energetic arguments \citep{bonnavion2022use}, due to the increase of the kinetic energy of large-scale structures shed in the wake, as highlighted in Figures \ref{fig:4POD_cumsum} and \ref{fig:URANS_vorticity}.
For strong boat-tailing actuation ($p\geq2$), the fluidic pinball behaves as a single bluff body with a globally unsteady wake, as seen in the modal analysis from Section \ref{subsec:POD}. This suggests the existence of an optimal range for the boat-tailing actuation, beyond which additional input energy does not further reduce drag and may even promote more intense unsteady dynamics.

Regarding measurement uncertainty, the width of the uncertainty envelope is highly affected by the presence of sporadic high-amplitude fluctuations in the load cell signals. This effect is particularly pronounced for strong base-bleeding actuation, for which the resulting uncertainty increases noticeably. Although the force statistical distributions remain approximately symmetric, the load cell signals are significantly affected by small perturbations. This increase in uncertainty is primarily associated with the actuation-induced load component. Because this component depends on the operating conditions at the time of each actuation case, it is more susceptible to external disturbances such as mechanical vibrations and environmental noise, which are unavoidable in the present experimental configuration. These effects are therefore reflected in the reported uncertainty band rather than removed through post-processing.


%% file: Figures/FP_WakeFlowDevelopment.tex
\begin{tikzpicture}

    \tikzset{
      weakAct/.style={
        ->,
        line width=0.6pt,
        black,
        >=Stealth
      },
      strongAct/.style={
        ->,
        line width=1pt,
        black,
        >=Stealth
      }
    }
    
    \node[anchor=south west,inner sep=0] (img) at (0,0)
        {\includegraphics{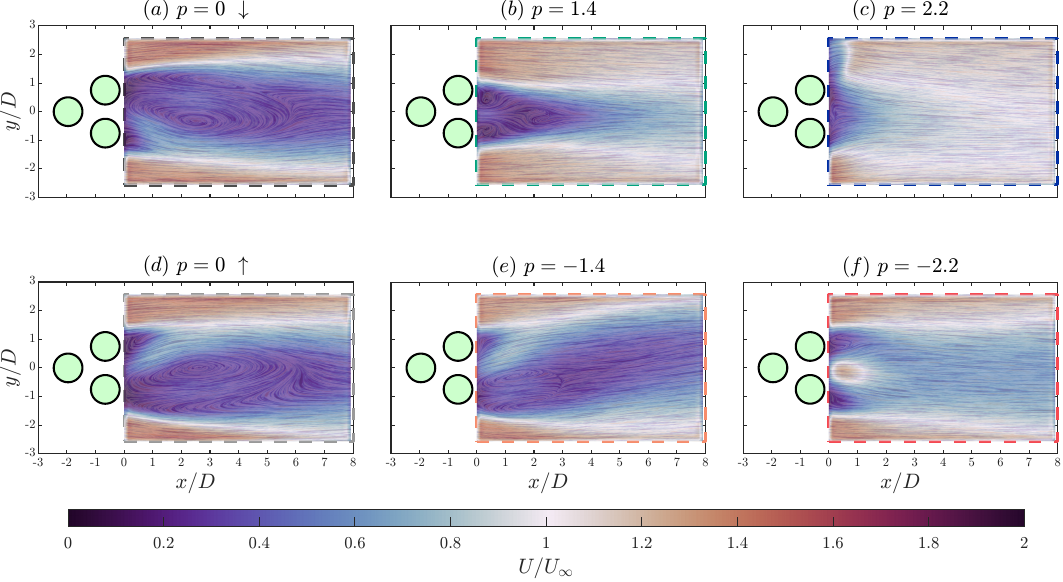}};

    \begin{scope}[x={(img.south east)},y={(img.north west)}]

    \def\D{10pt}   
    \def\Dlarge{6pt}   

    \coordinate (bC2) at (0.4313,0.8439);
    \coordinate (bC3) at (0.4313,0.7705);

    \draw[line width=0.8pt,bt14]
        (bC2) ++(150:\D) arc (150:90:\D);
    \path
      (bC2) ++(150:\D) coordinate (start)
      arc (150:90:\D) coordinate (end);
    \draw[-{Stealth[scale=0.75]}, line width=0.6pt,bt14]
    (end) -- ++(0:0.005);

    \draw[line width=0.8pt,bt14]
        (bC3) ++(-150:\D) arc (-150:-90:\D);
    \path
      (bC3) ++(-150:\D) coordinate (start)
      arc (-150:-90:\D) coordinate (end);
    \draw[-{Stealth[scale=0.75]}, line width=0.6pt,bt14]
    (end) -- ++(0:0.005);

    \coordinate (cC2) at (0.7623,0.8439);
    \coordinate (cC3) at (0.7623,0.7705);

    \draw[line width=1.2pt,bt22]
        (cC2) ++(150:\D) arc (150:50:\D);
    \path
      (cC2) ++(150:\D) coordinate (start)
      arc (150:50:\D) coordinate (end);
    \draw[-{Stealth[scale=0.85]}, line width=0.8pt,bt22]
    (end) -- ++(-50:0.01);

    \draw[line width=1.2pt,bt22]
        (cC3) ++(-150:\D) arc (-150:-50:\D);
    \path
      (cC3) ++(-150:\D) coordinate (start)
      arc (-150:-50:\D) coordinate (end);
    \draw[-{Stealth[scale=0.85]}, line width=0.8pt,bt22]
    (end) -- ++(50:0.01);

    \coordinate (eC2) at (0.4313,0.404);
    \coordinate (eC3) at (0.4313,0.3305);
    \draw[line width=0.8pt,bb14]
        (eC2) ++(90:\D) arc (90:145:\D);
    \path
      (eC2) ++(90:\D) coordinate (start)
      arc (90:145:\D) coordinate (end);
    \draw[-{Stealth[scale=0.75]}, line width=0.8pt,bb14]
    (end) -- ++(-115:0.01);

    \draw[line width=0.8pt,bb14]
        (eC3) ++(-90:\D) arc (-90:-145:\D);
    \path
      (eC3) ++(-90:\D) coordinate (start)
      arc (-90:-145:\D) coordinate (end);
    \draw[-{Stealth[scale=0.75]}, line width=0.8pt,bb14]
    (end) -- ++(115:0.01);

    \coordinate (fC2) at (0.7623,0.404);
    \coordinate (fC3) at (0.7623,0.3305);
    \draw[line width=1.2pt,bb22]
        (fC2) ++(45:\D) arc (45:145:\D);
    \path
      (fC2) ++(45:\D) coordinate (start)
      arc (45:145:\D) coordinate (end);
    \draw[-{Stealth[scale=0.85]}, line width=0.8pt,bb22]
    (end) -- ++(-115:0.01);

    \draw[line width=1.2pt,bb22]
        (fC3) ++(-45:\D) arc (-45:-145:\D);
    \path
      (fC3) ++(-45:\D) coordinate (start)
      arc (-45:-145:\D) coordinate (end);
    \draw[-{Stealth[scale=0.85]}, line width=0.8pt,bb22]
    (end) -- ++(115:0.01);

    \end{scope}
    \end{tikzpicture}

%% file: Figures/FP_POD_phi_all.tex
\begin{tikzpicture}

    \tikzset{
        weakAct/.style={
        ->,
        line width=0.48pt,
        black,
        >=Stealth
      },
        strongAct/.style={
        ->,
        line width=0.8pt,
        black,
        >=Stealth
      }
    }
    
    \node[anchor=south west,inner sep=0] (img) at (0,0)
        {\includegraphics{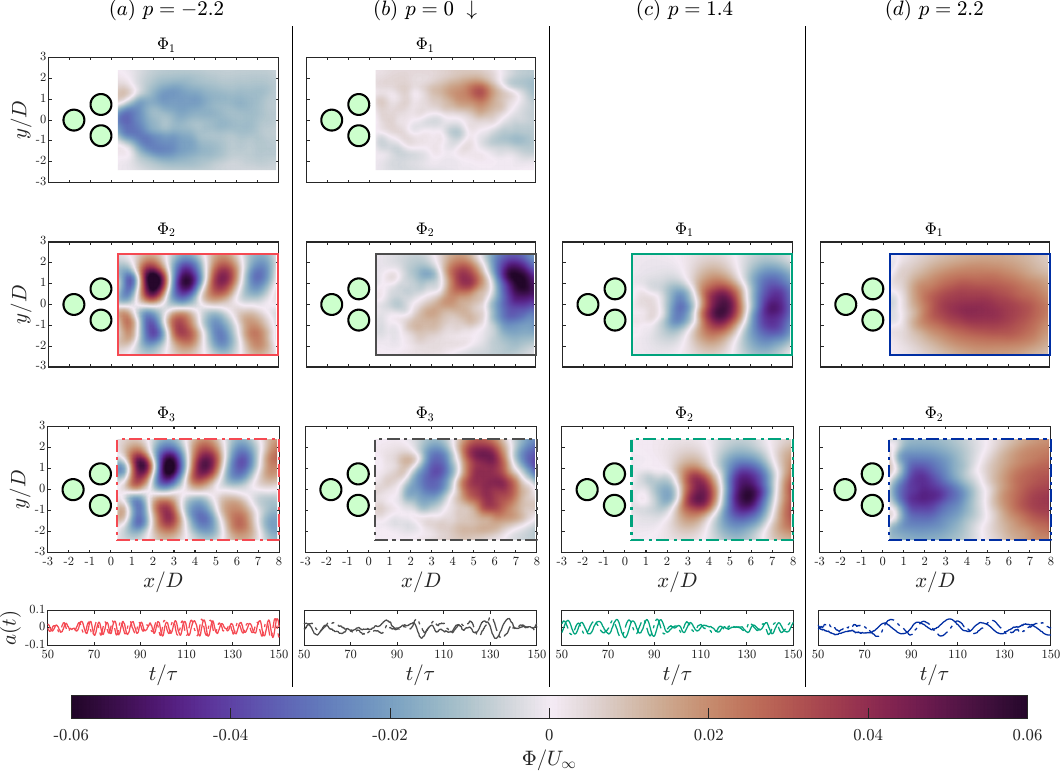}};

    \begin{scope}[x={(img.south east)},y={(img.north west)}]

    \def\D{8pt}   
    \def\Dlarge{4pt}   

    \coordinate (aC2) at (0.0954,0.8644);
    \coordinate (aC3) at (0.0954,0.824);
    \draw[line width=1.2pt,bb22]
        (aC2) ++(45:\D) arc (45:145:\D);
    \path
      (aC2) ++(45:\D) coordinate (start)
      arc (45:145:\D) coordinate (end);
    \draw[-{Stealth[scale=0.85]}, line width=0.8pt,bb22]
    (end) -- ++(-115:0.01);

    \draw[line width=1.2pt,bb22]
        (aC3) ++(-45:\D) arc (-45:-145:\D);
    \path
      (aC3) ++(-45:\D) coordinate (start)
      arc (-45:-145:\D) coordinate (end);
    \draw[-{Stealth[scale=0.85]}, line width=0.8pt,bb22]
    (end) -- ++(115:0.01);

    \coordinate (aC2) at (0.0954,0.626);
    \coordinate (aC3) at (0.0954,0.5856);
    \draw[line width=1.2pt,bb22]
        (aC2) ++(45:\D) arc (45:145:\D);
    \path
      (aC2) ++(45:\D) coordinate (start)
      arc (45:145:\D) coordinate (end);
    \draw[-{Stealth[scale=0.85]}, line width=0.8pt,bb22]
    (end) -- ++(-115:0.01);

    \draw[line width=1.2pt,bb22]
        (aC3) ++(-45:\D) arc (-45:-145:\D);
    \path
      (aC3) ++(-45:\D) coordinate (start)
      arc (-45:-145:\D) coordinate (end);
    \draw[-{Stealth[scale=0.85]}, line width=0.8pt,bb22]
    (end) -- ++(115:0.01);

    \coordinate (aC2) at (0.0954,0.3876);
    \coordinate (aC3) at (0.0954,0.3472);
    \draw[line width=1.2pt,black,bb22]
        (aC2) ++(45:\D) arc (45:145:\D);
    \path
      (aC2) ++(45:\D) coordinate (start)
      arc (45:145:\D) coordinate (end);
    \draw[-{Stealth[scale=0.85]}, line width=0.8pt,bb22]
    (end) -- ++(-115:0.01);

    \draw[line width=1.2pt,bb22]
        (aC3) ++(-45:\D) arc (-45:-145:\D);
    \path
      (aC3) ++(-45:\D) coordinate (start)
      arc (-45:-145:\D) coordinate (end);
    \draw[-{Stealth[scale=0.85]}, line width=0.8pt,bb22]
    (end) -- ++(115:0.01);

    \coordinate (cC2) at (0.579,0.626);
    \coordinate (cC3) at (0.579,0.5856);
    \draw[line width=0.8pt,bt14]
        (cC2) ++(150:\D) arc (150:90:\D);
    \path
      (cC2) ++(150:\D) coordinate (start)
      arc (150:90:\D) coordinate (end);
    \draw[-{Stealth[scale=0.75]}, line width=0.6pt,bt14]
    (end) -- ++(0:0.005);

    \draw[line width=0.8pt,bt14]
        (cC3) ++(-150:\D) arc (-150:-90:\D);
    \path
      (cC3) ++(-150:\D) coordinate (start)
      arc (-150:-90:\D) coordinate (end);
    \draw[-{Stealth[scale=0.75]}, line width=0.6pt,bt14]
    (end) -- ++(0:0.005);

    \coordinate (cC2) at (0.579,0.3876);
    \coordinate (cC3) at (0.579,0.3472);
    \draw[line width=0.8pt,bt14]
        (cC2) ++(150:\D) arc (150:90:\D);
    \path
      (cC2) ++(150:\D) coordinate (start)
      arc (150:90:\D) coordinate (end);
    \draw[-{Stealth[scale=0.75]}, line width=0.6pt,bt14]
    (end) -- ++(0:0.005);

    \draw[line width=0.8pt,bt14]
        (cC3) ++(-150:\D) arc (-150:-90:\D);
    \path
      (cC3) ++(-150:\D) coordinate (start)
      arc (-150:-90:\D) coordinate (end);
    \draw[-{Stealth[scale=0.75]}, line width=0.6pt,bt14]
    (end) -- ++(0:0.005);

    \coordinate (dC2) at (0.8212,0.626);
    \coordinate (dC3) at (0.8212,0.5856);
    \draw[line width=1.2pt,bt22]
        (dC2) ++(150:\D) arc (150:50:\D);
    \path
      (dC2) ++(150:\D) coordinate (start)
      arc (150:50:\D) coordinate (end);
    \draw[-{Stealth[scale=0.85]}, line width=0.8pt,bt22]
    (end) -- ++(-50:0.01);

    \draw[line width=1.2pt,bt22]
        (dC3) ++(-150:\D) arc (-150:-50:\D);
    \path
      (dC3) ++(-150:\D) coordinate (start)
      arc (-150:-50:\D) coordinate (end);
    \draw[-{Stealth[scale=0.85]}, line width=0.8pt,bt22]
    (end) -- ++(50:0.01);

    \coordinate (dC2) at (0.8212,0.3876);
    \coordinate (dC3) at (0.8212,0.3472);
        \draw[line width=1.2pt,bt22]
        (dC2) ++(150:\D) arc (150:50:\D);
    \path
      (dC2) ++(150:\D) coordinate (start)
      arc (150:50:\D) coordinate (end);
    \draw[-{Stealth[scale=0.85]}, line width=0.8pt,bt22]
    (end) -- ++(-50:0.01);

    \draw[line width=1.2pt,bt22]
        (dC3) ++(-150:\D) arc (-150:-50:\D);
    \path
      (dC3) ++(-150:\D) coordinate (start)
      arc (-150:-50:\D) coordinate (end);
    \draw[-{Stealth[scale=0.85]}, line width=0.8pt,bt22]
    (end) -- ++(50:0.01);

    \end{scope}
    \end{tikzpicture}

%% file: Figures/URANS_vorticity.tex
\begin{tikzpicture}

    \tikzset{
      weakAct/.style={
        ->,
        line width=0.6pt,
        black,
        >=Stealth
      },
      strongAct/.style={
        ->,
        line width=1pt,
        black,
        >=Stealth
      }
    }
    
    \node[anchor=south west,inner sep=0] (img) at (0,0)
        {    \includegraphics[scale=0.55, trim={0cm 4cm 0cm 0cm}, clip]{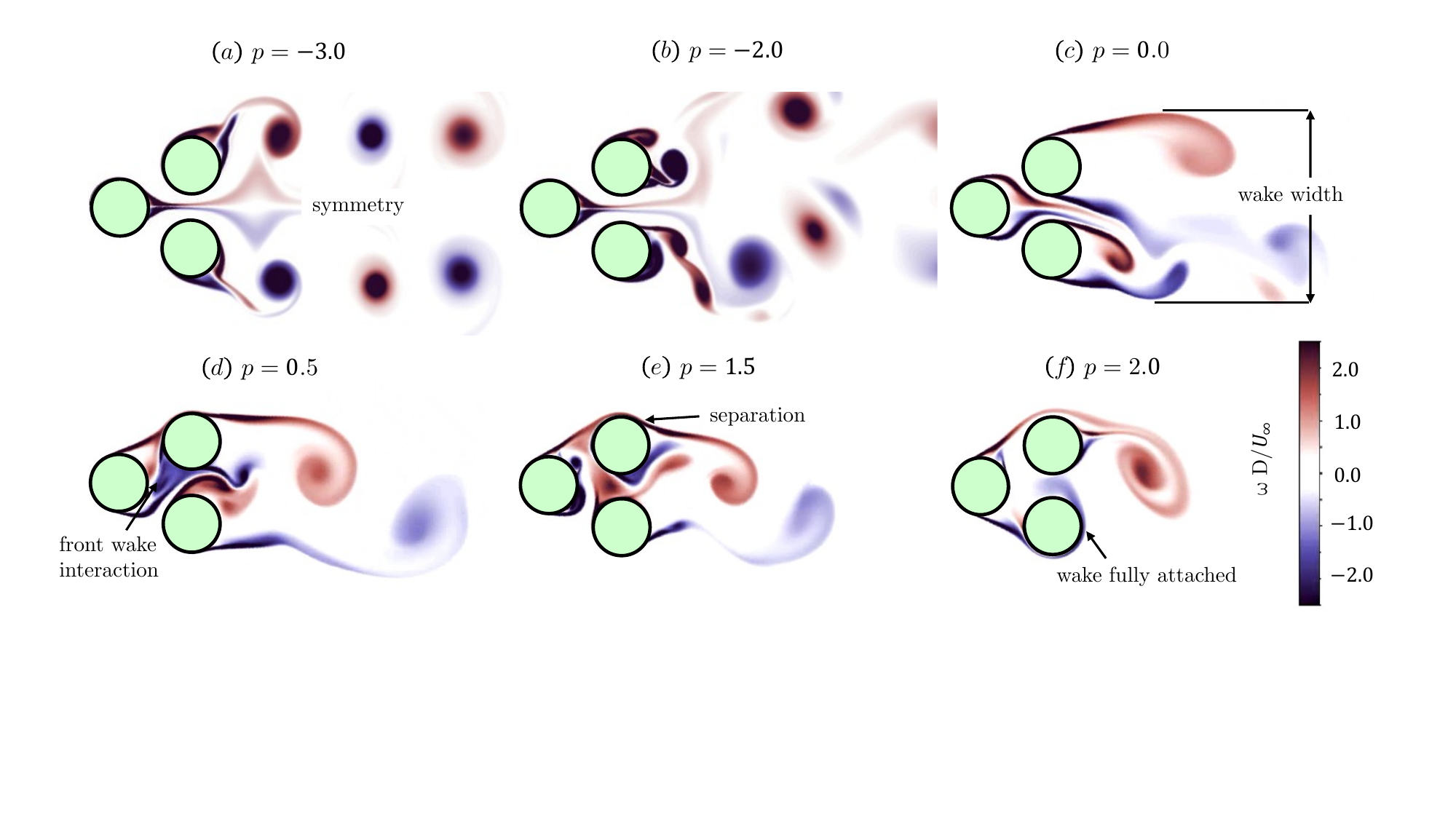}};

    \begin{scope}[x={(img.south east)},y={(img.north west)}]

    \def\D{7pt}   

    \coordinate (aC2) at (0.1317,0.7435);
    \coordinate (aC3) at (0.1317,0.616);

    \draw[line width=1.5pt,black]
        (aC2) ++(0:\D) arc (0:180:\D);
    \path
      (aC2) ++(0:\D) coordinate (start)
      arc (0:180:\D) coordinate (end);
    \draw[-{Stealth[scale=0.9]}, line width=1.2pt,black]
    (end) -- ++(-90:0.018);

    \draw[line width=1.5pt,black]
        (aC3) ++(-0:\D) arc (-0:-180:\D);
    \path
      (aC3) ++(-0:\D) coordinate (start)
      arc (-0:-180:\D) coordinate (end);
    \draw[-{Stealth[scale=0.9]}, line width=1.2pt,black]
    (end) -- ++(90:0.018);

    \coordinate (aC2) at (0.4275,0.742);
    \coordinate (aC3) at (0.4275,0.6145);

    \draw[line width=1.2pt,black]
        (aC2) ++(45:\D) arc (45:180:\D);
    \path
      (aC2) ++(45:\D) coordinate (start)
      arc (45:180:\D) coordinate (end);
    \draw[-{Stealth[scale=0.8]}, line width=1pt,black]
    (end) -- ++(-90:0.018);

    \draw[line width=1.2pt,black]
        (aC3) ++(-45:\D) arc (-45:-180:\D);
    \path
      (aC3) ++(-45:\D) coordinate (start)
      arc (-45:-180:\D) coordinate (end);
    \draw[-{Stealth[scale=0.8]}, line width=1pt,black]
    (end) -- ++(90:0.018);

    \coordinate (aC2) at (0.1317,0.32);
    \coordinate (aC3) at (0.1317,0.1925);

    \draw[line width=0.6pt,black]
        (aC2) ++(180:\D) arc (180:90:\D);
    \path
      (aC2) ++(180:\D) coordinate (start)
      arc (180:90:\D) coordinate (end);
    \draw[-{Stealth[scale=0.75]}, line width=0.5pt,black]
    (end) -- ++(-0:0.005);

    \draw[line width=0.6pt,black]
        (aC3) ++(-180:\D) arc (-180:-90:\D);
    \path
      (aC3) ++(-180:\D) coordinate (start)
      arc (-180:-90:\D) coordinate (end);
    \draw[-{Stealth[scale=0.75]}, line width=0.5pt,black]
    (end) -- ++(0:0.005);

    \coordinate (aC2) at (0.4275,0.312);
    \coordinate (aC3) at (0.4275,0.1845);

    \draw[line width=1pt,black]
        (aC2) ++(180:\D) arc (180:75:\D);
    \path
      (aC2) ++(180:\D) coordinate (start)
      arc (180:75:\D) coordinate (end);
    \draw[-{Stealth[scale=0.75]}, line width=0.9pt,black]
    (end) -- ++(-55:0.015);

    \draw[line width=1pt,black]
        (aC3) ++(-180:\D) arc (-180:-75:\D);
    \path
      (aC3) ++(-180:\D) coordinate (start)
      arc (-180:-75:\D) coordinate (end);
    \draw[-{Stealth[scale=0.75]}, line width=0.9pt,black]
    (end) -- ++(55:0.015);

    \coordinate (aC2) at (0.7233,0.312);
    \coordinate (aC3) at (0.7233,0.1845);

    \draw[line width=1.2pt,black]
        (aC2) ++(180:\D) arc (180:45:\D);
    \path
      (aC2) ++(180:\D) coordinate (start)
      arc (180:45:\D) coordinate (end);
    \draw[-{Stealth[scale=0.8]}, line width=1pt,black]
    (end) -- ++(-75:0.018);

    \draw[line width=1.2pt,black]
        (aC3) ++(-180:\D) arc (-180:-45:\D);
    \path
      (aC3) ++(-180:\D) coordinate (start)
      arc (-180:-45:\D) coordinate (end);
    \draw[-{Stealth[scale=0.8]}, line width=1pt,black]
    (end) -- ++(75:0.018);

    \end{scope}
\end{tikzpicture}

%% file: Chapters/4-ROM_discussion.tex
\section{Reduced-order interpretation of the wake dynamics}
\label{sec:ROM}

\begin{figure*}[!t]
\centering
\input{Figures/FP_ROM/FP_ROM}
\caption{Conceptual sketch of a reduced-order representation of the turbulent wake of the fluidic pinball. It provides a synthesis of the experimentally observed wake dynamics as the actuation parameter $p$ is varied. The evolution of wake deflection, drag coefficient, and dominant oscillatory \ac{POD} modes is shown, highlighting symmetry-breaking and symmetrization transitions, as well as the loss of control authority at strong actuation. Insets illustrate mean wake structures (represented as contours of the velocity magnitude, with superposed \ac{LIC} representation of the mean velocity field) and modal phase portraits associated with each regime.}
\label{fig:5ROM_figure}
\end{figure*}
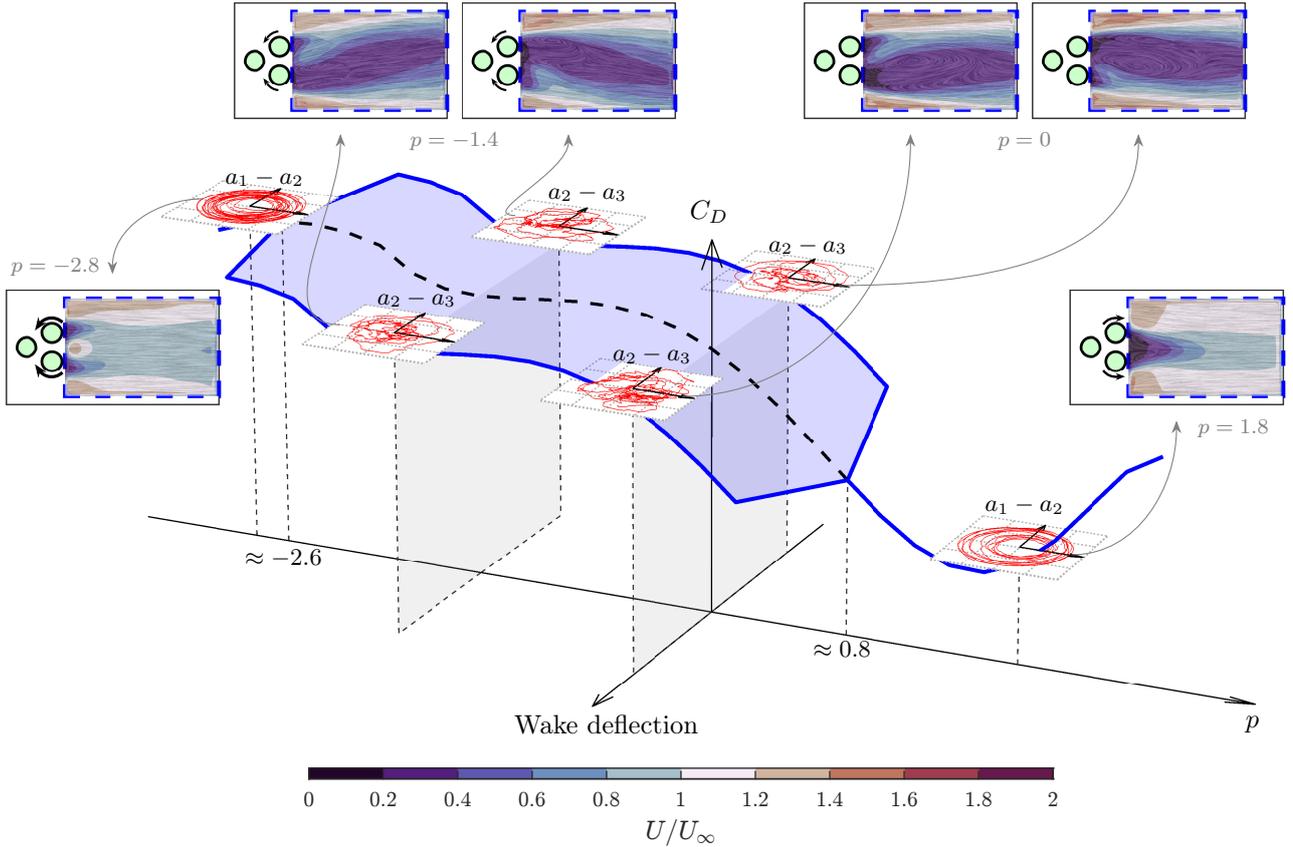

The purpose of this section is to interpret the wake dynamics and force response of the turbulent fluidic pinball within a low-dimensional framework.

The fluidic pinball configuration has emerged as a canonical testbed for the development and validation of \acp{ROM}, due to its rich and structured wake dynamics. In particular, low-order models explored at low Reynolds numbers have demonstrated that the flow evolves in a low-dimensional space and undergoes successive bifurcations as control parameters are varied \citep{deng2020low}. In those studies, the Reynolds number was the primary bifurcation parameter, and the resulting \ac{ROM} captured the transition between different wake regimes. However, such investigations were limited to laminar flow, with $Re\leq 130$.

In the present experiments, the fluidic pinball is studied in a turbulent regime under active control. Although turbulence introduces broadband fluctuations and higher-dimensional dynamics, the results presented above indicate that the dominant wake behavior and force response remain governed by a small number of physically meaningful degrees of freedom. In this context, the actuation parameter $p$ replaces the Reynolds number as the relevant control parameter, driving transitions between asymmetric, symmetric, and bluff-body-like wake states.

While the present work does not aim to derive a quantitative predictive \ac{ROM}, the experimental results provide strong evidence that the dominant wake dynamics evolve on a low-dimensional space parameterized by the actuation input. Specifically, the combined force measurements, modal analysis, and wake topology indicate that the essential physics of the controlled turbulent wake can be represented using three independent coordinates. This reduced representation provides a compact and interpretable description of the flow, without requiring the full time series of velocity fields or aerodynamic forces. 

Figure \ref{fig:5ROM_figure} provides a schematic summary of this reduced-order interpretation, illustrating how changes in actuation drive transitions between wake asymmetry, wake symmetrization, and bluff-body behavior.

The fluidic pinball undergoes a discrete bifurcation associated with wake symmetrization under symmetric actuation. Although the flow might be expected to remain symmetric in the absence of actuation, it is well known that the unforced wake of the fluidic pinball selects one of two asymmetric states depending on the initial conditions, with no spontaneous switching unless external inputs are applied \citep{lam1988phenomena}. The application of symmetric control stabilizes the wake between the two rear cylinders. Complete wake symmetrization is observed for  $p\geq0.8$ in the case of boat tailing, and for $p\leq-2.6$ in the case of base bleeding. 

These observations are fully consistent with the modal analysis. In asymmetric wake configurations, the leading POD mode captures wake deflection, while the dominant oscillatory dynamics appear in the second and third modes. Conversely, for symmetric wake states, the oscillatory vortex-shedding dynamics emerge in the first two POD modes.

Within this reduced-order framework, the evolution of the drag coefficient can be interpreted as a consequence of state-space transitions, rather than as an isolated force response. Drag reduction is evidently visible once the wake collapses onto the symmetric branch of the reduced space, whereas the subsequent drag increase at strong actuation reflects a transition toward a bluff-body-dominated space. In contrast, base-bleeding actuation produces a continuous and gradual increase in drag across the explored parameter range.


The available data suggest that a three-dimensional state space is sufficient to represent the dominant physics of the controlled fluidic pinball wake. This space may be spanned by a wake-deflection mode, a drag-related mode, and the actuation parameter, providing a compact yet physically meaningful foundation for future reduced-order modeling efforts. From an experimental point of view, this reduced-order interpretation provides a unifying architecture to connect force measurements, modal content, and wake topology, offering guidance for control design and comparison across Reynolds numbers without requiring full-order experiments or simulations.

%% file: Figures/FP_ROM/FP_ROM.tex
\begin{tikzpicture}
\def\D{0.2} 

\node (base) {\includegraphics{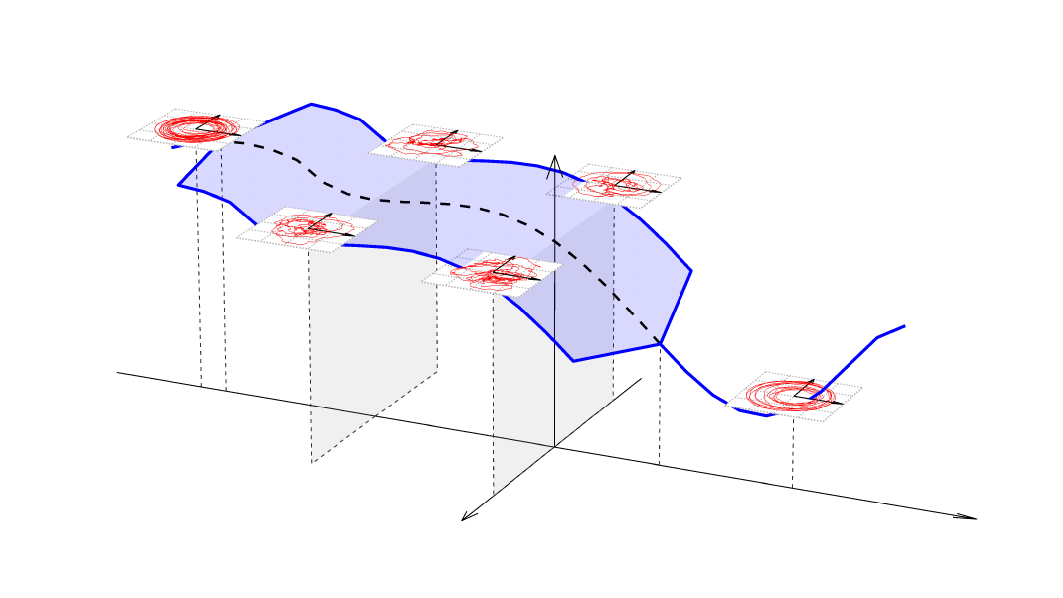}};


\node 
  at ([xshift=-1cm,yshift=-4cm]base.center)
  {Wake deflection};

\node
  at ([xshift=7.5cm,yshift=-4cm]base.center)
  {$p$};

\node
  at ([xshift=0.35cm,yshift=2.8cm]base.center)
  {$C_D$};

\node[font=\small]
  at ([xshift=4.5cm,yshift=-1.1cm]base.center)
  {$a_1-a_2$};

\node[font=\small]
  at ([xshift=-5.5cm,yshift=3.25cm]base.center)
  {$a_1-a_2$};

\node[font=\small]
  at ([xshift=-1.25cm,yshift=3.05cm]base.center)
  {$a_2-a_3$};
\node[font=\small]
  at ([xshift=-3.5cm,yshift=1.65cm]base.center)
  {$a_2-a_3$};
  
\node[font=\small]
  at ([xshift=1.65cm,yshift=2.35cm]base.center)
  {$a_2-a_3$};
\node[font=\small]
  at ([xshift=-0.4cm,yshift=0.9cm]base.center)
  {$a_2-a_3$};

\node[font=\small]
  at ([xshift=2.1cm,yshift=-3cm]base.center)
  {$\approx0.8$};
\node[font=\small]
  at ([xshift=-5.25cm,yshift=-1.8cm]base.center)
  {$\approx-2.6$};


\node (fp0d)
  at ([xshift=6cm,yshift=4.8cm]base.center)
  {\includegraphics{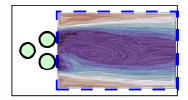}};

\node[font=\footnotesize,gray]
  at ([xshift=-1.5cm,yshift=-0.1cm]fp0d.south)
  {$p=0$};

\draw[->, gray,  >=Stealth]
  (2.05,1.82)
  to[out=0,in=-90]
  (fp0d.south);

\node (fp0u)
  at ([xshift=3cm,yshift=4.8cm]base.center)
  {\includegraphics{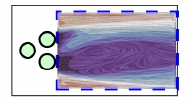}};


\draw[->,gray,>=Stealth,thin]
  (0,0.35)
  to[out=0,in=-90]
  (fp0u.south);

\node (fp14d)
  at ([xshift=-1.5cm,yshift=4.8cm]base.center)
  {\includegraphics{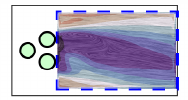}};

\node[font=\footnotesize,gray]
  at ([xshift=-1.5cm,yshift=-0.1cm]fp14d.south)
  {$p=-1.4$};

\draw[->, gray,  >=Stealth]
  (-2.2,2.75)
  to[out=180,in=-90]
  (fp14d.south);
  
\coordinate (aC2) at (-2.3,4.99);
\coordinate (aC3) at (-2.3,4.61);

\draw[line width=0.8pt,black]
    (aC2) ++(95:\D) arc (95:145:\D);
\path
  (aC2) ++(95:\D) coordinate (start)
  arc (95:145:\D) coordinate (end);
\draw[-{Stealth[scale=0.5]}, line width=0.6pt,black]
(end) -- ++(-125:0.09);

\draw[line width=0.8pt,black]
    (aC3) ++(-95:\D) arc (-95:-145:\D);
\path
  (aC3) ++(-95:\D) coordinate (start)
  arc (-95:-145:\D) coordinate (end);
\draw[-{Stealth[scale=0.5]}, line width=0.6pt,black]
(end) -- ++(125:0.09);

\node (fp14u)
  at ([xshift=-4.5cm,yshift=4.8cm]base.center)
  {\includegraphics{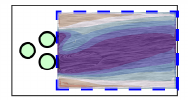}};


\draw[->,gray,>=Stealth,thin]
  (-4.5,1.3)
  to[out=180,in=-90]
  (fp14u.south);

\coordinate (aC2) at (-5.3,4.99);
\coordinate (aC3) at (-5.3,4.61);

\draw[line width=0.8pt,black]
    (aC2) ++(95:\D) arc (95:145:\D);
\path
  (aC2) ++(95:\D) coordinate (start)
  arc (95:145:\D) coordinate (end);
\draw[-{Stealth[scale=0.5]}, line width=0.6pt,black]
(end) -- ++(-125:0.09);

\draw[line width=0.8pt,black]
    (aC3) ++(-95:\D) arc (-95:-145:\D);
\path
  (aC3) ++(-95:\D) coordinate (start)
  arc (-95:-145:\D) coordinate (end);
\draw[-{Stealth[scale=0.5]}, line width=0.6pt,black]
(end) -- ++(125:0.09);

\node (fp28)
  at ([xshift=-7.5cm,yshift=1cm]base.center)
  {\includegraphics{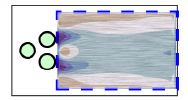}};

\node[font=\footnotesize, gray]
  at ([xshift=-0.75cm,yshift=0.1cm]fp28.north) {$p=-2.8$};

\draw[->, gray, >=Stealth]
  (-6.3,2.97)
  to[out=180,in=90]
  (fp28.north);

\coordinate (aC2) at (-8.3,1.19);
\coordinate (aC3) at (-8.3,0.81);

\draw[line width=1.2pt,black]
    (aC2) ++(45:\D) arc (45:145:\D);
\path
  (aC2) ++(45:\D) coordinate (start)
  arc (45:145:\D) coordinate (end);
\draw[-{Stealth[scale=0.75]}, line width=0.8pt,black]
(end) -- ++(-125:0.1);

\draw[line width=1.2pt,black]
    (aC3) ++(-45:\D) arc (-45:-145:\D);
\path
  (aC3) ++(-45:\D) coordinate (start)
  arc (-45:-145:\D) coordinate (end);
\draw[-{Stealth[scale=0.75]}, line width=0.8pt,black]
(end) -- ++(125:0.1);

\node (fp18)
  at ([xshift=6.5cm,yshift=1cm]base.center)
  {\includegraphics{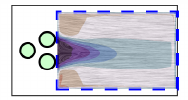}};

\node[font=\footnotesize, gray]
  at ([xshift=0.75cm,yshift=-0.1cm]fp18.south) {$p=1.8$};

\draw[->, gray, >=Stealth]
  (5.12,-1.75)
  to[out=0,in=-90]
  (fp18.south);

\coordinate (aC2) at (5.7,1.19);
\coordinate (aC3) at (5.7,0.81);

\draw[line width=0.9pt,black]
    (aC2) ++(145:\D) arc (145:80:\D);
\path
  (aC2) ++(145:\D) coordinate (start)
  arc (145:80:\D) coordinate (end);
\draw[-{Stealth[scale=0.6]}, line width=0.65pt,black]
(end) -- ++(0:0.09);

\draw[line width=0.9pt,black]
    (aC3) ++(-145:\D) arc (-145:-80:\D);
\path
  (aC3) ++(-145:\D) coordinate (start)
  arc (-145:-80:\D) coordinate (end);
\draw[-{Stealth[scale=0.6]}, line width=0.65pt,black]
(end) -- ++(0:0.09);

\node (fpcolorbar)
  at ([xshift=0cm,yshift=-5.1cm]base.center)
  {\includegraphics{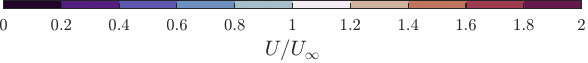}};


\end{tikzpicture}

%% file: Chapters/5-Conclusions.tex
\section{Conclusions}
\label{sec:conclusions}

The flow past the fluidic pinball was studied experimentally at $Re=9100$. Time-resolved velocity field measurements were combined with synchronized force measurements to characterize the wake dynamics and the resulting aerodynamic loads under controlled rotation of the cylinders. The experiment was focused on steady symmetric actuation cases, in which the upstream cylinder remained stationary while the two downstream cylinders rotated with equal and opposite angular velocities. 

In the baseline configuration without actuation, the wake exhibited a strong recirculation region and a deflected gap jet between the two downstream cylinders. The orientation of this gap jet depended on the initial conditions and external perturbations, leading to two quasi-stable asymmetric wake states. No spontaneous switching between these states was observed; however, the wake orientation could be modified through external actuation, confirming the inherent sensitivity of the configuration.

Under base-bleeding actuation, outward rotation of the downstream cylinders strengthened the gap jet and led to a gradual increase in drag. Velocity field measurements revealed a persistent splitting of the wake, consistent with the elevated drag levels. Wake symmetrization was only achieved at sufficiently strong actuation ($p=-2.6$), at which a sudden decrease in drag was observed. \ac{POD} supported this interpretation, asymmetric cases were characterized by a leading non-oscillatory mode associated with gap-jet deflection, whereas symmetric wakes exhibited the dominant oscillatory vortex-shedding dynamics in the leading mode pair.

In contrast, inward rotation of the downstream cylinders in the boat-tailing configuration produced a pronounced reduction in drag. Wake symmetrization occurred at moderate actuation strengths, preceding the minimum drag, which was attained at $p\approx1.8$. Beyond this optimal value, further increases in actuation strength resulted in a loss of control authority: the drag coefficient increased again despite the wake remaining symmetric. In this regime, the fluidic pinball behaved as a single bluff body with globally unsteady dynamics. The strong rotation altered the boundary-layer development on the upstream cylinder, diminishing the effectiveness of the control mechanism and promoting energetic unsteady structures.

These results indicate that the dominant flow physics of the fluidic pinball at turbulent Reynolds numbers under symmetric actuation can be described in a low-dimensional space. The actuation parameter $p$ acts as a bifurcation parameter governing transitions between asymmetric and symmetric wake states and defining an optimal range for the boat-tailing parameter. A discrete wake-deflection coordinate captures the orientation of the gap jet, while the drag coefficient provides a quantitative measure of performance and loss of control authority. These remarks provide a qualitative foundation for the development of \acp{ROM} of the turbulent fluidic pinball and highlight the distinction between mean-flow stabilization and the suppression of unsteady wake dynamics.

The present findings show that the fluidic pinball 
constitutes a geometrically simple prototypical configuration that features the complete set of dynamics of more complex bluff-bodies in the turbulent regime. 
Boat tailing, for instance, is a generic flow control technique to mitigate vortex shedding and to reduce drag for two- and three-dimensional bluff bodies \citep{Hucho2011book}.
Pitchfork bifurcations have been reported for the drag crisis of the circular cylinder flow \citep{Schewe1983jfm} as a function of Reynolds number
and Ahmed body wakes \citep{Grandemange2013jfm,Barros2017jfm}
as a function of the gap clearance.

%% file: Chapters/A-Uncertainty.tex
\section{Force coefficient uncertainty estimation}
\label{app1}

This appendix details the procedure used to estimate the uncertainty of the drag force coefficient. The force coefficient can be defined as:
\begin{equation}
    C_D =  \frac{F_x}{\frac{1}{2}\rho U_\infty^2 DH},
\end{equation}
where  $F_x=F_{x_\mathrm{baseline}}+F_{x_\mathrm{act}}$. Here, $F_{x_\mathrm{baseline}}$ is the baseline load and $F_{x_\mathrm{act}}$ is the actuation-induced load, which are independent. The uncertainties of the velocity and force measurements contribute to the overall uncertainty of $C_D$. 

The forces are obtained from the load cell signals through the inverse calibration matrix $\mathsfbi{K}^{-1}$. Since only the streamwise component is required, the corresponding row of $\mathsfbi{K}^{-1}$ is denoted by $\mathbf{g}_x$. The drag force is therefore computed directly from the measured load differences as:

\begin{equation}
F_{x_\mathrm{baseline}}
= \mathbf{g}_x
\left(
\mathbf{L}_{\mathrm{WT\ on}}
-
\mathbf{L}_{\mathrm{WT\ off}}
\right),
\end{equation}
\begin{equation}
F_{x_\mathrm{act}}
= \mathbf{g}_x
\left(
\mathbf{L}_{\mathrm{WT\ on,\ act\ on}}
-
\mathbf{L}_{\mathrm{WT\ on,\ act\ off}}
\right).
\end{equation}

Applying the standard error propagation formula \citep{moffat1988describing}, under the assumption of independent measurements, gives:
\begin{equation}
    \left(\frac{\Delta C_D}{C_D}\right)^2 = \left(\frac{\Delta F_x}{F_x}\right)^2 + \left(\frac{2\Delta U_\infty}{U_\infty}\right)^2.
\end{equation}

The velocity uncertainty $\Delta U_\infty$ comes directly from the PIV measurements as described in the main text. The remaining terms concern the force measurements.

The total force uncertainty in the streamwise direction, under the assumption of independent terms, is defined as:
\begin{equation}
    \Delta F_x^2 = \Delta F_{x_\mathrm{baseline}}^2 + \Delta F_{x_\mathrm{act}}^2 + \Delta K_x^2 + \Delta L^2,
\end{equation}
where $\Delta F_{x_\mathrm{baseline}}$ and $\Delta F_{x_\mathrm{act}}$ are the statistical contributions of the baseline and actuation-induced load components, respectively, $\Delta K_x$ is the uncertainty associated with the calibration matrix for the streamwise component, and $\Delta L$ is the load-cell specification uncertainty.

The manufacturer provides hysteresis, linearity, and repeatability errors. These combine independently as:
\begin{equation}
    \Delta L^2=L_\mathrm{nom}^2(\Delta_\mathrm{hys}^2+\Delta_\mathrm{lin}^2+\Delta_\mathrm{rep}^2),
\end{equation}
where $L_\mathrm{nom}$ is the nominal load.

The uncertainty contribution from the calibration matrix in the streamwise direction is written as:
\begin{equation}
    \Delta K_x^2 = F_x^2\Delta_\mathrm{cal}^2,
\end{equation}
where $\Delta_\mathrm{cal}$ is the calibration error specified by the manufacturer.

The resulting statistical contribution from the actuation-induced force is:
\begin{equation}
    \Delta F_{x_\mathrm{act}}^2=\frac{1}{N_\mathrm{eff}}\mathbf{g}_x\bm{\Sigma}_{L_\mathrm{act}}\mathbf{g}_x^\top,
\end{equation}
where $\bm{\Sigma}_{L_\mathrm{act}}$ is the covariance matrix of the differenced signal from the instantaneous load cell signals and $N_\mathrm{eff}$ is the effective number of samples. Because the load-cell time series are temporally correlated, the effective number of samples is:
\begin{equation}
    N_{\mathrm{eff},j}=\frac{N}{1+2\sum_{k=1}^{N-1}\left(1-\frac{k}{N}\right)\rho_{k_j}},
\end{equation}
where $\rho_{k_j}$ is the autocorrelation of the $j$-th load cell signal. The overall effective sample size is taken as:
\begin{equation}
    N_\mathrm{eff}=\min_j N_{\mathrm{eff},j}.
\end{equation}

The baseline force uncertainty is treated analogously, and therefore:
\begin{equation}
\begin{split}
    \Delta F_{x_\mathrm{baseline}}^2 &= \frac{1}{{N_\mathrm{eff}}}\mathbf{g}_x\bm{\Sigma}_{L_\mathrm{baseline}}\mathbf{g}_x^\top,\\     
\end{split}
\end{equation}
where $\bm{\Sigma}_{L_\mathrm{baseline}}$ is the covariance matrix of the differenced baseline signal. The same definition of $N_\mathrm{eff}$ applies.

Overall, the force coefficient uncertainty can be expressed as:
\begin{equation}
    \frac{\Delta C_D}{C_D}=\sqrt{\frac{\Delta F_{x_\mathrm{baseline}}^2 + \Delta F_{x_\mathrm{act}}^2 + \Delta K_x^2 + \Delta L^2}{F_x^2} + \frac{4\Delta U_\infty^2}{U_\infty^2}}.
\end{equation}

Specifically, for the baseline case ($p=0$), the relative uncertainty is evaluated using representative values obtained from the experimental measurements. The numerical values reported below correspond to force-related uncertainties expressed in millinewtons and velocity uncertainty expressed in millimeters per second.

Note that using
\[
F_x = \SI{2.2}{\newton}, \quad
\Delta F_{x_\mathrm{baseline}} = \SI{11}{\milli\newton}, \quad
\Delta F_{x_\mathrm{act}} = \SI{37}{\milli\newton},
\]
\[
\Delta K_x = \SI{2.2}{\milli\newton}, \quad
\Delta L = \SI{4.3}{\milli\newton}, 
\]
\[
U_\infty = \SI{0.31}{\metre\per\second}, \quad
\Delta U_\infty = \SI{0.26}{\milli\metre\per\second},
\]
the relative drag-coefficient uncertainty becomes:
\begin{equation}
    \frac{\Delta C_D}{C_D}=\sqrt{\frac{11^2 + 37^2 + 2.2^2 + 4.3^2}{2.2^2} + \frac{4(0.26)^2}{
    0.31^2}}\times10^{-3}
\end{equation}
which yields
\[
\frac{\Delta C_D}{C_D} = 1.8 \times 10^{-2}.
\]

This value is representative of the average relative uncertainty reported for the drag coefficient across all actuation cases.